\documentclass[%
 aip,
 jcp,%
 amsmath,amssymb,
 preprint,%
]{revtex4-1}
 
\usepackage{graphicx}
\usepackage{dcolumn}
\usepackage{bm}
\usepackage{braket}
\usepackage{dsfont}
\usepackage{placeins}
\usepackage{nicefrac}
\usepackage{natbib} 
\usepackage{float}
\usepackage[usenames,dvipsnames]{color}

\usepackage{xr}
\usepackage{amsmath}
\usepackage[boxruled]{algorithm2e}

\begin{document}
\preprint{AIP/123-QED}
 
\title[]{Girsanov reweighting for path ensembles and Markov state models}

\author{L. Donati}
 \affiliation{Department of Biology, Chemistry, Pharmacy, Freie Universit\"at Berlin, Takustra\ss e 3, D-14195 Berlin, Germany}

\author{C. Hartmann}
\affiliation{Institute of Mathematics, Brandenburgische Technische Universit\"at Cottbus-Senftenberg,
Konrad-Wachsmann-Allee 1, D-03046 Cottbus, Germany}
 
\author{B.G. Keller}
 \email{bettina.keller@fu-berlin.de}
\affiliation{Department of Biology, Chemistry, Pharmacy, Freie Universit\"at Berlin, Takustra\ss e 3, D-14195 Berlin, Germany}

\date{\today}

\begin{abstract}
The sensitivity of molecular dynamics on changes in the potential energy function plays an important role in understanding the dynamics and function of 
complex molecules.
We present a method to obtain path ensemble averages of a perturbed dynamics from a 
set of paths generated by a reference dynamics. 
It is based on the concept of path probability measure and the Girsanov theorem, a result from stochastic analysis to estimate a change of measure of a path ensemble. 
Since Markov state models (MSM) of the molecular dynamics can be formulated as a combined phase-space and path ensemble average, the method can be extended to
reweight MSMs by combining it with a reweighting of the Boltzmann distribution. 
We demonstrate how to efficiently implement the Girsanov reweighting in a molecular dynamics simulation program by calculating parts of the 
reweighting factor ``on the fly'' during the simulation, and we benchmark the method on test systems ranging from a two-dimensional diffusion process to an artificial many-body system and alanine dipeptide and valine dipeptide in implicit and explicit water.
The method can be used to study the sensitivity of molecular dynamics on external perturbations as well as to reweight trajectories generated by enhanced sampling schemes to the original dynamics.
\end{abstract}


\maketitle

\section{Introduction}

\begin{figure*}[h]
  \begin{center}
  \includegraphics[scale=1]{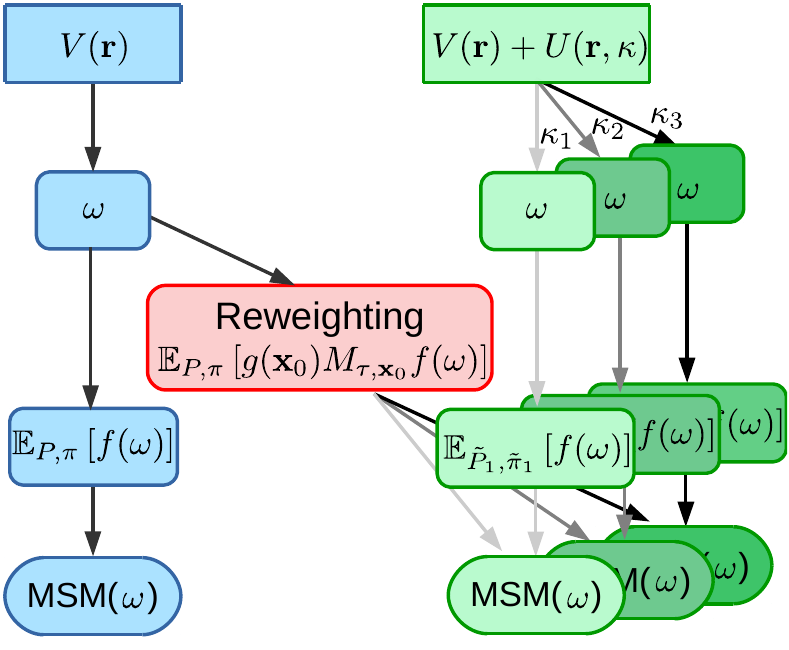}
    \caption{Workflow of a reweighting scheme. $V(\mathbf{r})$ is the reference potential energy function, $V(\mathbf{r}) + U(\mathbf{r},\kappa)$ is the perturbed potential energy function, where $\kappa$ is a tunable parameter, $\omega$ is the trajectory generated by an MD simulation at the potential $V(\mathbf{r})$ or $V(\mathbf{r}) + U(\mathbf{r},\kappa)$ and at a fixed thermodynamic state point. A MSM is an expectation value with respect to a path probability measure $P$ and a stationary distribution $\pi$, which can be estimated from the trajectory $\omega$. A dynamical reweighting scheme, reweights the path probability measures $P$ and $\pi$ of the reference dynamics, to the path probability measures of a perturbed potential energy functions $\widetilde P$ and $\widetilde \pi$. Thus, we can use the trajectory generated at $V(\mathbf{r})$ to estimate dynamical expectation value (e.g. MSMs) at the perturbed potential energy functions. 
    }
    \label{fig:intro}
  \end{center}
\end{figure*}

Molecular dynamics (MD) simulations with explicit solvent molecules are routinely used as efficient importance sampling algorithms for the Boltzmann distribution of molecular systems. 
From the conformational snapshots created by MD simulations, one can estimate phase-space ensemble averages and thus interpret experimental data or thermodynamic functions in terms of molecular conformations.
In recent years, the scope of MD simulations has considerably widened, and the method has been increasingly used to construct models of the conformational dynamics
\cite{Bolhuis2002, Faradjian2004, Best2005, Hegger2009, Faccioli2010}.
Most notably Markov state models \cite{Schuette1999, Schuette1999b,  Deuflhard2000, Swope2004, Chodera2007, Buchete2008, Prinz2011, Keller2010, Keller2011}, in which the conformational space is discretized into disjoint states and the dynamics is modeled as a Markov jump process between these states, have become a valuable tool for the analysis of complex molecular dynamics \cite{Voelz2010b, Stanley2014, Bowman2015, Plattner2015, Zhang2016}.
For the construction of dynamic models, one has to estimate path ensemble averages. 
For example in MSMs, the transition probabilities between a pair of states $B_i$ and $B_j$ is estimated by considering a set of paths $S_{\tau} = \lbrace \omega_1, \omega_2, ... \omega_n \rbrace$, each of which has length $\tau$, counting the number of paths which start in $B_i$ and end in $B_j$, and comparing this number to the total number of paths in the set
(vertical line of blue boxes in Fig.~\ref{fig:intro}).
%

\begin{figure*}[h]
  \begin{center}
  \includegraphics[scale=1]{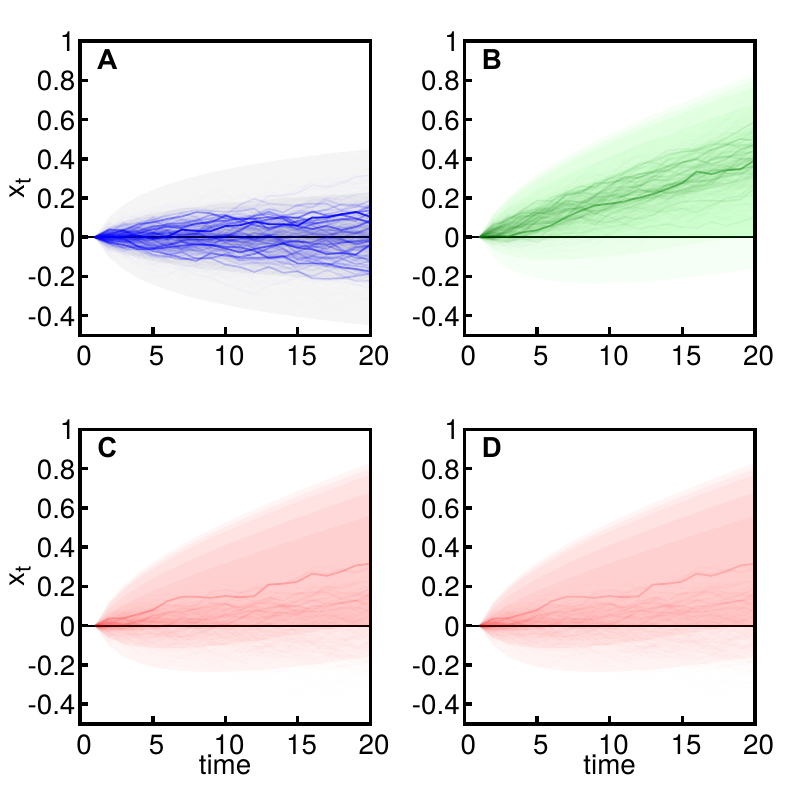}
    \caption{Two sets of trajectories starting from $x(t=0)=0$.
   	Set $S_{\tau}$ generated by a  Brownian motion without drift, associated to path probability measure $P$.
   	Set $\widetilde S_{\tau}$ generated by a  Brownian motion with drift, associated to path probability measure $\widetilde P$.
	(A) Set $S_{\tau}$, color intensity represents $P$.
	(B) Set $\widetilde S_{\tau}$, color intensity represents $\widetilde P$.
	(C) Set $S_{\tau}$, color intensity represents $\widetilde P$ (direct calculation).
	(D) Set $S_{\tau}$, color intensity represents $\widetilde P$ (Girsanov formula).}
    \label{graph:path_ensemble}
    \end{center}
\end{figure*}

Suppose, one would like to compare the dynamics in a reference potential energy function $V(\mathbf{r})$ to the dynamics in a series of perturbed potential energy functions $V(\mathbf{r}) + U(\mathbf{r}, \kappa)$, where $U(\mathbf{r}, \kappa)$ represents the perturbation and $\kappa$ is a tunable parameter, e.g.~a force constant.
While for phase-space ensemble averages, numerous methods exist to reweight the samples of the reference conformational ensemble to yield ensemble averages for the perturbed systems \cite{Zwanzig1954, Bennett1976, Kumar1992, Kaestner2005}, similar reweighting schemes have not yet been developed for path ensemble averages
of explicit-solvent simulations.
This means that currently one would have to re-simulate the dynamics at each parameter value (vertical lines of green boxes in Fig.~\ref{fig:intro})  
and then construct a MSM for each simulation separately.  
This is computationally extremely costly.
An alternative would be to reweight the path ensemble average at the reference potential energy function to obtain path ensemble averages for the perturbed systems  (reweighting box in Fig.~\ref{fig:intro}). 
From measure theory it is well known that a reweighting factor is given as the ratio between the probability measure associated to the reference potential energy function and the probability measure associated to the perturbed potential energy function. 
This applies to reweighting phase space ensemble averages as well as to reweighting path ensemble averages. 
Fig.~\ref{graph:path_ensemble} illustrates the idea of a path ensemble reweighting method. 
The figure shows two sets of paths, one generated by a Brownian dynamics simulation without drift, $S_{\tau}$ (Fig.~\ref{graph:path_ensemble}.A), 
and one generated by a Brownian dynamics simulation with drift, $\widetilde S_{\tau}$ (Fig.~\ref{graph:path_ensemble}.B).
Both simulations sample the same path space $\Omega_{\tau, \mathbf{x}}$ but the probability with which a given path is realized differs in the two simulations.
In Fig.~\ref{graph:path_ensemble}.A and \ref{graph:path_ensemble}.B, the sets of paths, $S_{\tau}$ and $\widetilde S_{\tau}$, are colored according to their respective path probability density $\mu_{P}(\omega)$ and $\mu_{\widetilde P}(\omega)$.
Fig.~\ref{graph:path_ensemble}.C and \ref{graph:path_ensemble}.D shows again $S_{\tau}$, this time however we colored the paths according to the probability density $\mu_{\widetilde P}(\omega)$ with which they would have been generated by a Brownian dynamics with drift.
For Brownian dynamics this probability can be calculated directly (Fig.~\ref{graph:path_ensemble}.C), for other types of dynamics a reweighting method has to be used 
(Fig.~\ref{graph:path_ensemble}.D).
To estimate a path ensemble average for the Brownian dynamics with drift from $S_{\tau}$, the contribution of each path $\omega$ to the estimate is multiplied by the ratio 
$M_{\tau, \mathbf{x}}(\omega) = \mu_{\widetilde P}(\omega)/ \mu_{P}(\omega)$
(Fig.~\ref{fig:intro}).
%

Path ensemble reweighting schemes have initially been developed in the field of importance sampling for stochastic differential equations \cite{LPS98, PR09, HS12}.
For Langevin dynamics, the Girsanov theorem \cite{Girsanov1960, Oksendal2003} provides us with an expression for the probability ratio, and thus reweighting path ensemble averages becomes possible for this type of dynamics (section \ref{sec:dynamicalReweighting}). 
It has recently been demonstrated that the theorem can be applied to reweight Markov state models of Brownian dynamics in  one- and two-dimensional potential energy functions
 \cite{Schuette2015b}.
Here we demonstrate how the Girsanov reweighting scheme can be applied to explicit-solvent all-atom MD simulations. For this, we need to address
to critical pillars on which the Girsanov reweighting scheme rests:
\begin{itemize}
	\item The equation of motion need to contain a stochastic term which generates random forces drawn from a normal distribution (white noise).
	\item To calculate the reweighting factor, the random forces need to be accessible for each MD simulation step.
\end{itemize}	
In all-atom MD simulations, the system is propagated by the Newton equations of motion which do not contain a stochastic term. We will discuss, how the Girsanov theorem can nonetheless be applied to this type of simulations
(section \ref{sec:stochasticForces}). 
The second point, in principle, requires that the forces are written out at every MD simulation step, i.e. at a frequency of femtoseconds rather than the usual output rate of several picoseconds. This quickly fills up any hard disc and slows the simulation by orders of magnitudes.
We will present a computational efficient implementation of the reweighting scheme (section \ref{sec:implementation}).
When applying the Girsanov theorem to reweight an MSM, an additional difficulty arises
\begin{itemize}
	\item The degrees of freedom which are affected by the perturbation might not be part of the relevant subspace of the MSM.
\end{itemize}
We found that the reweighting becomes problematic in this case and propose to project the perturbed degrees of freedom onto the relevant subspace during the estimation of 
the ratio of probability measures (section \ref{sec:projection}).
The Girsanov reweighting method is demonstrated and benchmarked on several systems, ranging from two-dimensional diffusion processes (section \ref{sec:twodimsys}), 
over molecular model systems which follow a Langevin dynamics (section \ref{sec:manybody})
to all-atom MD simulations of alanine and valine dipeptides in explicit and implicit solvent (section \ref{sec:alanineValine}).

\section{Theory}
\subsection{Molecular dynamics}
Consider a molecular system with $N$ particles, which evolves in time $t$ according to the Langevin equation
\begin{eqnarray}
\label{eq:unperturbed} 
M \frac{ \mathrm{d} \mathbf{v}(t)}{\mathrm{d}t}&=& -\nabla V\left(\mathbf{r}(t)\right) - \gamma \mathbf{v}(t) + \sigma \eta(t)  \cr
\mathbf{v}(t) &=&  \frac{ \mathrm{d} \mathbf{r}(t)}{\mathrm{d} t} \, ,
\end{eqnarray}
where  $M$ is the mass matrix, $\mathbf{r}(t)$ and $\mathbf{v}(t) \in \mathbb{R}^{3 N}$ are the position vector and the velocity vector.
$V(\mathbf{r})$ is the potential energy function.
The interaction with the thermal bath is modelled by the friction coefficient $\gamma$, and an uncorrelated Gaussian white noise $\eta(t) \in \mathbb{R}^{3 N}$  which is scaled by the volatility $\sigma$ 
\begin{equation}
\sigma = \sqrt{2 k_B T \gamma M}
\label{eq:EinsteinRelation}
\end{equation}
where $k_B$ is the Boltzmann constant and $T$ is the temperature of the system. 
%

%
The phase-space vector $\mathbf{x}(t)=\lbrace \mathbf{r}(t), \mathbf{v}(t) \rbrace \in \Gamma$ fully represents the state of the system at time $t$, 
where $\Gamma=\mathbb{R}^{6N}$ denotes the phase space of the system.
The dynamics in eq.~\ref{eq:unperturbed} is associated to an equilibrium probability density 
\begin{equation}
	\mu_\pi(\mathbf{x}) 
	= \frac{\exp\left[-\beta \mathcal{H}(\mathbf{x})\right]}{Z}
\label{eq:stationaryDist}	
\end{equation}
where $\beta=\frac{1}{k_B T}$,  $\mathcal{H}(\mathbf{x})=  \frac{1}{2} \mathbf{v}^\top M \mathbf{v} + V(\mathbf{r})$ is the classical Hamiltonian of the system, 
and $Z = \int_{\Gamma} \exp\left[-\beta \mathcal{H}(\mathbf{x})\right] \,\mathrm{d}\mathbf{x}$ is the partition function. 
The function $\mu_\pi(\mathbf{x})$ is associated to the probability measure
\begin{equation}
	\pi(A) = \mathbb{P}(\mathbf{x}\in A) =  \int_A \mu_\pi (\mathbf{x}) \mathrm{d} \mathbf{x} \ , \quad \forall A \subset \Gamma
\label{eq:phaseSpaceMeasure}	
\end{equation}
where $\pi(A)$ represents the equilibrium probability of finding the system in a subset $A$ of the phase space $\Gamma$.
The expectation value of a function $a(\mathbf{x}): \Gamma \rightarrow \mathbb{R}$ with respect to the probability density $\mu_{\pi}(\mathbf{x})$ is given as
\begin{eqnarray}
	\mathbb{E}_\pi\left[ a\right]	
	&=& \int_{\Gamma} a(\mathbf{x}) \mu_\pi(\mathbf{x}) \,\mathrm{d}\mathbf{x} 
	=	\lim_{n\rightarrow \infty} \frac{1}{n} \sum_{\mathbf{x}_k\in S_n} a(\mathbf{x}_k)\, ,
\label{eq:phaseSpaceAverage}	
\end{eqnarray}
where $S_n = \lbrace \mathbf{x}_1, ... \mathbf{x}_n\rbrace$ is a set of states distributed according to eq.~\ref{eq:stationaryDist}. 
When the phase space vectors $\mathbf{x}_i$ are generated by numerically integrating eq.~\ref{eq:unperturbed}, the second equality only holds if the sampling is ergodic.
Eq.~\ref{eq:phaseSpaceAverage} defines a \emph{phase-space ensemble average}. 
The subscript $\pi$ indicates the measure for which the expectation value is calculated.

\subsection{Path ensembles and MSMs}

A path $\omega=\lbrace\mathbf{x}(t=0)=\mathbf{x}_0,\mathbf{x}_1,\mathbf{x}_2,...,\mathbf{x}(\tau)=\mathbf{x}_n\rbrace$ is a time-discretized realization of the dynamics $\mathbf{x}(t)$ on the time interval $[0,\tau = n\cdot \Delta t]$ starting at particular point $\mathbf{x}_0 \in \Gamma$, where $\Delta t$ is the time step and $n \in \mathbb{N}$ is the number of time steps.
The associated path space is denoted $\Omega_{\tau,\mathbf{x}} = \mathbb{R}^{6N \cdot n}$.
A subset of the path space $\mathcal{A}$ is constructed as a product of subsets $A_i \subset \Gamma$ of the state space
$\mathcal{A}=A_1 \times A_2 ... \times A_n$, where the subset $A_i$ represents the phase space volume in which $\mathbf{x}_i$
may be found.
The probability that by integrating eq.~\ref{eq:unperturbed} one obtains a path $\omega$ which belongs to the subset $\mathcal{A} \subset \Omega_{\tau,\mathbf{x}}$
is given as 
\begin{eqnarray}
\label{eq:path_measure1}
P(\mathcal{A})
	&= &  \mathbb{P}(\omega \in \mathcal{A}) = \mathbb{P}(\mathbf{x}_1 \in A_1,\mathbf{x}_2 \in A_2,...,\mathbf{x}_\tau \in A_n) \cr
%
	& = & \int_{A_1}\int_{A_2}...\int_{A_n} 
			p(\mathbf{x}_0,\mathbf{x}_1; \, \Delta t) \, p(\mathbf{x}_1,\mathbf{x}_2; \, \Delta t)\, ... \, p(\mathbf{x}_{n-1},\mathbf{x}_n;\, \Delta t) 	  \, \mathrm{d}\mathbf{x}_1 \, \mathrm{d}\mathbf{x}_2 \, ... \, \mathrm{d}\mathbf{x}_n \, . 
\end{eqnarray}
The function $p(\mathbf{x}_i,\mathbf{x}_{i+1}; \, \Delta t)$ is the transition probability density , 
i.e. the conditional probability to be in $\mathbf{x}_{i+1}$ after a time $\Delta t$ given the initial state $\mathbf{x}_i$.
The function $P$ is a path probability measure and is the analogon to $\pi$ in phase space ensemble averages (eq.~\ref{eq:phaseSpaceMeasure}).
The path probability measure is associated to the path probability density function:
\begin{eqnarray}
\label{eq:path_probabilityDensity1}
	 \mu_{P}(\omega) 
	 &=& \mu_{P}(\mathbf{x}_1,\mathbf{x}_2,...,\mathbf{x}_{\tau})  
	= 	p(\mathbf{x}_0,\mathbf{x}_1; \, \Delta t) \, p(\mathbf{x}_1,\mathbf{x}_2; \, \Delta t)\, ... \, p(\mathbf{x}_{n-1},\mathbf{x}_{n};\, \Delta t)
\end{eqnarray}
and hence the formal analogon to eq.~\ref{eq:phaseSpaceMeasure} in the path space is
\begin{eqnarray}
	P({\mathcal{A}})  	
	&=& \mathbb{P}(\omega \in \mathcal{A}) = \int_{\mathcal{A}}  \mu_{P}(\omega) \mathrm{d}Ê\omega
	\ , \quad \forall \mathcal{A} \subset \Omega_{\tau, \mathbf{x}}
\end{eqnarray}
where the integration over $\mathrm{d}\omega$ is defined by eq.~\ref{eq:path_measure1}.

Let $f: \Omega_{\tau,\mathbf{x}} \rightarrow \mathbb{R}$ be an integrable function, 
which assigns a real number to each path.
The expectation value of this function is
\begin{eqnarray}
	\mathbb{E}_{P}[f(\omega)]	 
	&=&	\int_{\Omega_{\tau, \mathbf{x}}}	f(\omega) \, \mu_{P}(\omega)\,  \mathrm{d}\omega \cr
	&=& \int_{\Gamma} \int_{\Gamma} \dots  \int_{\Gamma} 	
		 f(\mathbf{x}_1,\mathbf{x}_2,...,\mathbf{x}_n) \mu_{P}(\mathbf{x}_1,\mathbf{x}_2,...,\mathbf{x}_n)
		 \mathrm{d}\mathbf{x}_1 \, \mathrm{d}\mathbf{x}_2 \, ... \, \mathrm{d}\mathbf{x}_n\cr
	&=& \lim_{m\rightarrow \infty} \frac{1}{m} \sum_{\omega_k \in S_{\tau,\mathbf{x},m}} f(\omega_k) \, .
\label{eq:pathEnsembleAverage}	
\end{eqnarray} 
where we again assumed that the paths have a common initial state $\mathbf{x}(t=0) = \mathbf{x}_0$, 
and $S_{\tau, \mathbf{x},m} = \lbrace \omega_1, \omega_2, ... \omega_mÊ\rbrace$ corresponds to a set of paths of length $\tau$ generated by numerically integrating eq.~\ref{eq:unperturbed}.
When the paths are extracted from a single long trajectory, the last equality only holds  if the sampling is ergodic.
Eq.~\ref{eq:pathEnsembleAverage} defines a \emph{path ensemble average}.
The subscript $P$ indicates that the expectation value is calculated with respect to a path probability measure.

For Markov processes, one can define a transition probability density 
$p(\mathbf{x}, \mathbf{y}; \tau)$, i.e.  the conditional probability to be in $\mathbf{x}_n=\mathbf{y}$ after a time $\tau$ given that the path started in $\mathbf{x}_0= \mathbf{x}$,
by integrating the path probability density over all intervening states and applying recursively the Chapman-Kolmogorov equation
\begin{eqnarray}
 	p(\mathbf{x}, \mathbf{y}, \tau) 
	= 	\int_{\Gamma} \int_{\Gamma} \dots  \int_{\Gamma} \mu_{P}(\mathbf{x}_1,\mathbf{x}_2,...,\mathbf{y}) \, 
		\mathrm{d}\mathbf{x}_1 \, \mathrm{d}\mathbf{x}_2 \, ... \, \mathrm{d}\mathbf{x}_{n-1} \, .
\label{eq:transitionProbabilityDensity01}
\end{eqnarray}
%
%

%
Markov processes can be approximated by Markov state models 
\cite{Schuette1999, Schuette1999b,  Deuflhard2000,  Swope2004, Chodera2007, Buchete2008, Prinz2011, Keller2011}.
In these models, the phase space is discretized into disjoint sets (or microstates) ${B_1, B_2, ... B_s}$ with $\cup_{i=1}^m B_i = \Gamma$,
where the indicator function of the $i$th state is given by 
\begin{equation}
\mathbf{1}_{B_i}(\mathbf{x}) :=
\begin{cases}
1 &\text{if } \mathbf{x} \in B_i, \\
0 &\mbox{ otherwise\, .}
\label{eq:indicatorFunction}	
\end{cases}
\end{equation}
The associated cross-correlation function is
\begin{eqnarray}
	C_{ij}(\tau) 
	&=& \int_{\Gamma} \mu_{\pi} (\mathbf{x}) \mathbf{1}_{B_i}(\mathbf{x}) 
	\int_{\Gamma}	 p(\mathbf{x},\mathbf{y};\tau) \, \mathbf{1}_{B_j}(\mathbf{y}) \,  
	 \mathrm{d} \mathbf{y}\, \mathrm{d}\mathbf{x} \, ,
\label{eq:correlationMatrix}	
\end{eqnarray}
and the transition probability between set $B_i$ and set $B_j$ is  
\begin{eqnarray}
	T_{ij}(\tau) = \frac{C_{ij}(\tau)}{\sum_{j=1}^s  C_{ij}(\tau)} \, . 
\label{eq:transitionMatrix}	
\end{eqnarray}
$T_{ij}(\tau)$ are the elements of the transition matrix whose dominant eigenvectors and eigenvalues represent the slow dynamic processes of the system 
\cite{Schuette1999, Schuette1999b, Deuflhard2000, Swope2004, Chodera2007, Buchete2008, Prinz2011, Keller2011}. 
Because of eq.~\ref{eq:transitionProbabilityDensity01}, one can regard the inner integral of the cross-correlation function $C_{ij}(\tau)$ 
(eq.~\ref{eq:correlationMatrix}) as a path ensemble average (eq.~\ref{eq:pathEnsembleAverage})
which depends on the initial state $\mathbf{x}_0 = \mathbf{x}$ of the path ensemble
\begin{eqnarray}
	&&	\int_{\Gamma}	 p(\mathbf{x},\mathbf{y};\tau ) \, \mathbf{1}_{B_j}(\mathbf{y})\,  \mathrm{d} \mathbf{y} \cr
	&=&  \int_{\Gamma} \int_{\Gamma} \dots  \int_{\Gamma} \mu_{P}(\mathbf{x}_1,\mathbf{x}_2,...,\mathbf{x}_{n}) \, 
		\mathbf{1}_{B_j}(\mathbf{x}_n)\, 
		\mathrm{d}\mathbf{x}_1 \, \mathrm{d}\mathbf{x}_2 \, \dots \, \mathrm{d} \mathbf{x}_n \cr
	&=& 	\mathbb{E}_{P}^{\mathbf{x}_0}[\mathbf{1}_{B_j}(\mathbf{x}_n)]	 \, .
\end{eqnarray}
This expected value is a linear operator and it is called backward transfer operator (more details are given in appendix). The outer integral in eq.~\ref{eq:correlationMatrix} is a phase space ensemble average, thus we have
\begin{eqnarray}
	C_{ij}(\tau) 
	&=& \int_{\Gamma} \mu_{\pi} (\mathbf{x}_0)Ê\, \mathbf{1}_{B_i}(\mathbf{x}_0) \, 
		\mathbb{E}_{P}^{\mathbf{x}_0}[\mathbf{1}_{B_j}(\mathbf{x}_n)]  \mathrm{d}\mathbf{x}_0 \cr
	&=& 	\mathbb{E}_{\pi}[\mathbf{1}_{B_i}(\mathbf{x}_0) \, \mathbb{E}_{P}^{\mathbf{x}_0}[\mathbf{1}_{B_j}(\mathbf{x}_n)]] \cr
	&=&	\mathbb{E}_{ P, \pi}[\mathbf{1}_{B_i}(\mathbf{x}_0)\mathbf{1}_{B_j}(\mathbf{x}_n)]	
\end{eqnarray}
where the combined phase space and ensemble average is defined as 
\begin{eqnarray}
	\mathbb{E}_{P, \pi}[f(\omega)]	 
	&=& \int_{\Gamma} 
		\mu_{ \pi}(\mathbf{x}_0)	\int_{\Omega_{\tau, \mathbf{x}}} f(\omega) \, \mu_{P}(\omega)\,  \mathrm{d}\omega \, \mathrm{d}\mathbf{x}_0 \, .
\label{eq:pathEnsembleAverage02}		
\end{eqnarray}
Eq.~\ref{eq:pathEnsembleAverage02} extends eqs.~\ref{eq:pathEnsembleAverage}  to path ensembles with arbitrary initial states.
The elements $C_{ij}(\tau)$ can be estimated from a set of paths of length $\tau$, $S_{\tau,m}=\lbrace \nu_1, \nu_2 ... \nu_m \rbrace$, 
in which the initial states are no longer fixed but are distributed according to $\mu_{\pi}(\mathbf{x})$, 
\begin{eqnarray}
	C_{ij}(\tau) 
	&=& \lim_{mÊ\rightarrow \infty }\frac{1}{m}	\sum_{\nu_k \in  S_{\tau}}  \mathbf{1}_{B_i}( [\nu_k]_{t=0}) \cdot \mathbf{1}_{B_j}([\nu_k]_{t = n}) \cr
	&=& \lim_{mÊ\rightarrow \infty }\frac{1}{m}	\sum_{k=1}^m  \mathbf{1}_{B_i}([\mathbf{x}_0]_k) \cdot \mathbf{1}_{B_j}([\mathbf{x}_n]_k) 
\end{eqnarray}
where $ [\nu_k]_{t=i} = [\mathbf{x}_i]_k$ denotes the $i$th time step of the $k$th path.
%

\subsection{Dynamical reweighting}
\label{sec:dynamicalReweighting}
We alter the reference dynamics (eq.~\ref{eq:unperturbed}) by adding a perturbation $U(\mathbf{r}): \Gamma \rightarrow \mathbb{R}$
to the potential energy function $V(\mathbf{r})$. 
Thus, $\widetilde{V}(\mathbf{r}) = V(\mathbf{r}) + U(\mathbf{r})$ is the perturbed potential energy function and the Langevin equations of motion are
\begin{eqnarray}
\label{eq:perturbed_langevin}
M \frac{ \mathrm{d} \mathbf{v}(t)}{\mathrm{d} t}&=& -\nabla \widetilde{V}\left(\mathbf{r}(t)\right) - \gamma \mathbf{v}(t) + \sigma \eta(t)  \cr
\mathbf{v}(t) &=&  \frac{\mathrm{d} \mathbf{r}(t)}{\mathrm{d} t} \, ,
\end{eqnarray}
The perturbed dynamics is associated to a peturbed stationary probability density
\begin{equation}
\mu_{\widetilde\pi}(\mathbf{x}) 
= \frac{\exp\left[-\beta \widetilde{\mathcal{H}}(\mathbf{x})\right]}{\widetilde{Z}} \, ,
\end{equation}
where $\widetilde{\mathcal{H}}(\mathbf{x})=  \frac{1}{2} \mathbf{v}^\top M \mathbf{v} + \widetilde{V}(\mathbf{r})$
is the Hamiltonian of the perturbed system, and $\widetilde{Z}$ is its partition function.
The perturbation of the potential energy function also changes the transition probability density $\widetilde p(\mathbf{x}_i,\mathbf{x}_{i+1}; \, \Delta t)$, which gives rise to a
perturbed path probability density $\mu_{\widetilde P}(\omega)$ (eq.~\ref{eq:path_probabilityDensity1}) 
and a perturbed path measure $\widetilde{P}$  (eq.~\ref{eq:path_measure1}) .
Reweighting methods compare the probability measure of the perturbed systems to the probability measure of a reference system.
We first review the derivation of a reweighting scheme for phase space probability measures before discussing 
path space probability measures. 
The perturbed phase-space probability measure $\widetilde\pi$ is said to be \emph{absolutely continuous} with respect to the reference phase space probability measure $\pi$ if
\begin{eqnarray}
\widetilde \pi(A) = \int_A \mu_{\widetilde\pi}(\mathbf{x}) \, \mathrm{d}\mathbf{x} =0 \ &\Rightarrow& \ \pi(A) = \int_A \mu_\pi(\mathbf{x}) \, \mathrm{d}\mathbf{x} = 0  \, .
\label{eq:absoluteContinuity01}
\end{eqnarray}
This condition is sufficient and necessary to define the likelihood ratio between probability measures
\begin{equation}
g(\mathbf{x}) = \frac{\mathrm{d}\widetilde\pi}{\mathrm{d}\pi} = \frac{\mu_{\widetilde\pi}(\mathbf{x})}{\mu_\pi (\mathbf{x})} = \frac{Z}{\widetilde{Z}}\exp\left(-\beta U(\mathbf{x}) \right)\, .
\label{eq:RadonNikodym01}
\end{equation}
The function $g(\mathbf{x})$ is also called Radon-Nikodym derivative (Radon-Nikodym theorem \cite{Oksendal2003}) and can be used to construct the phase-space probability measure of the perturbed system, from the phase-space probability of the reference system:
\begin{equation}
\widetilde \pi(A) = \int_A \mu_{\widetilde\pi}(\mathbf{x}) \mathrm{d}\mathbf{x} = \int_A g(\mathbf{x}) \mu_\pi(\mathbf{x}) \mathrm{d}\mathbf{x} \, .
\label{eq:reweightPhaseSpace02}	
\end{equation}
As consequence, if $g(\mathbf{x})$ can be calculated, one can estimate a phase space ensemble average (eq.~\ref{eq:phaseSpaceAverage}) for the perturbed dynamics 
(eq.~\ref{eq:perturbed_langevin})
from a set of states $S_n = \lbrace \mathbf{x}_1, ... \mathbf{x}_n\rbrace$ which has been generated by the reference dynamics 
(eq.~\ref{eq:unperturbed})
\begin{eqnarray}
	\mathbb{E}_{\widetilde\pi}\left[ a\right]	
	&=& \int_{\Gamma} a(\mathbf{x}) \mu_{\widetilde\pi}(\mathbf{x}) \,\mathrm{d}\mathbf{x} 
	=	\int_{\Gamma} a(\mathbf{x}) g(\mathbf{x}) \mu_\pi(\mathbf{x}) \,\mathrm{d}\mathbf{x} 	\cr
	&=&	\lim_{n\rightarrow \infty} \frac{1}{n} \sum_{k=1}^n a(\mathbf{x}_k) g(\mathbf{x}_k)\, .
\label{eq:reweightPhaseSpace03}	
\end{eqnarray}
The notion of absolute continuity is valid also for path probability densities:
\begin{eqnarray}
\widetilde P(\mathcal{A}) = \int_\mathcal{A}  \mu_{\widetilde P}(\omega)  \mathrm{d}\omega=0 &\Rightarrow& 
P(\mathcal{A}) = \int_\mathcal{A} \mu_{P}(\omega) \mathrm{d}\omega= 0 \qquad
\forall \mathcal{A} \subset \Omega_{\tau,\mathbf{x}}\, .
\label{eq:absContPath}
\end{eqnarray}
Thus we can reweight a path ensemble average, by using the likelihood ratio between the path probability density $\mu_{\widetilde P}(\omega)$ and $\mu_{P}(\omega)$. For diffusion processes like \eqref{eq:unperturbed} and \eqref{eq:perturbed_langevin}, the likelihood ratio is given as  
\begin{equation}
M_{\tau, \mathbf{x}}(\omega)
= \frac{\mu_{\widetilde{P}}(\omega)}{\mu_P (\omega)} = \exp \left\lbrace - \sum_{i=1}^{3N} \left[ \sum_{k=0}^n \frac{\nabla_i U(\mathbf{r}_k)}{\sigma} \eta^i_k \sqrt{\Delta t} - \frac{1}{2} \sum_{k=0}^n  \left(\frac{\nabla_i U(\mathbf{r}_k)}{\sigma} \right)^2 \Delta t\right] \right\rbrace
\label{eq:reweightPathSpace01}
\end{equation}
where $\eta^i_k$ are the random numbers, along dimension $i$ at time step $k$, generated to integrate equation \ref{eq:unperturbed} of the reference dynamics
and $\nabla_i U(\mathbf{r}_k)$ is the gradient of the perturbation along dimension $i$ measured a the position $\mathbf{r}_k$. 
Note that to evaluate eq.~\ref{eq:reweightPathSpace01} one needs the positions and the random numbers for every time step of the time-discretized trajectory.
Eq.~\ref{eq:reweightPathSpace01}  is derived in appendix \ref{sec:GirsanovFormula}.
We remark that the quantity $M_{\tau, \mathbf{x}}(\omega)$ exists also for continuous paths ($\Delta t \rightarrow 0$). In this case the existence of the Radon-Nikodym derivative is guaranteed by the Girsanov theorem\cite{Girsanov1960, Oksendal2003} that states the conditions under which a perturbed path probability density $\widetilde P$ can be defined with respect to a reference path probability density $P$. 
The differences between time-continuous and time-discrete paths are discussed in the appendix \ref{sec:continuousPaths}.
Analogous to the reweighting of phase-space ensemble averages (eq.~\ref{eq:reweightPhaseSpace03}), 
we can use $M_{\tau, \mathbf{x}}(\omega)$ to reweight path ensemble averages
\begin{equation}
\begin{aligned}
	{\mathbb{E}}_{\widetilde P}[f(\omega)]	
	= \int_{\Omega_{\tau, \mathbf{x}}} 
		f(\omega) \, M_{\tau, \mathbf{x}}(\omega) \,  \mu_{P}(\omega) \, \mathrm{d}\omega \ 
	= 	 \lim_{n\rightarrow \infty} \frac{1}{n} \sum_{i=1}^n  M_{\tau, \mathbf{x}}(\omega_i) \, f(\omega_i) \, .	
	\end{aligned}
\label{eq:reweightPathEnsemble01}	
\end{equation}
The second equality shows how to estimate the path ensemble average $\mathbb{E}_{\widetilde P}[f(\omega)]$ at the perturbed dynamics (eq.~\ref{eq:perturbed_langevin})
from a set of paths $S_{\tau, \mathbf{x}, m} = \lbrace \omega_1, \omega_2, ... \omega_m \rbrace$ which has been generated by the reference dynamics 
(eq.~\ref{eq:unperturbed}).
%

\begin{table}
\centering
\caption{Overview of the notation}
\label{my-label}
    \begin{tabular}{| >{\centering\arraybackslash}m{1.8in} | >{\centering\arraybackslash}m{1.9in} | >{\centering\arraybackslash}m{2.1in} |>{\centering\arraybackslash}m{0.0000001in} }
    \cline{1-3}
                   & State: $\mathbf{x}$                                                                                           & Path:  $\omega$                                                                                                                                                                                                                                                                              &  \\[5ex] \cline{1-3}
Space              & $\Gamma\subset \mathbf{R}^{6N}$                                                                                                      & $\Omega_{\tau,\mathbf{x}} = \Gamma^n \subset \mathbf{R}^{6N\cdot n}$                                                                                                                                                                                                                                                                                   &  \\ [5ex]\cline{1-3}
Subsets            & $A_i \subset \Gamma$                                                                                          & $\mathcal{A} = \left( \prod_{i=1}^n A_i \right)\subset \Omega_{\tau,\mathbf{x}}$                                                                                                                                                                                                                                                                             &  \\ [5ex]\cline{1-3}
Probability density            & $\mu_\pi  (\mathbf{x}) \;|_{t\rightarrow \infty}$                                                                 & $\mu_{P}(\omega)$                                                                                                                                                                                                                                                                                                                          &  \\ [5ex]\cline{1-3}
Probability of a state/path      & $\pi(A)=\int_{A}\mu_\pi(\mathbf{x})\mathrm{d}\mathbf{x}$                                         & $P(\mathcal{A})=\int_{\mathcal{A}}\mu_{P}(\omega)\mathrm{d}\mathbf{\omega}$                                                                                                                                                                                                                             &  \\[5ex] \cline{1-3}
Expectation value  & $\mathbb{E}_\pi[f(\mathbf{x})] = \int_{A}f(\mathbf{x})\mu_\pi(\mathbf{x})\mathrm{d}\mathbf{x}$                & $\mathbb{E}_{P}[f(\omega)] = \int_{\mathcal{A}}f(\omega)\mu_{P}(\omega)\mathrm{d}\omega$                                                                                                                                                                                                                           &  \\ [5ex] \cline{1-3}
Absolute continuity         & $\widetilde \pi(A)=0 \Rightarrow \pi(A)=0$         & $\widetilde P(\mathcal{A}) =0\Rightarrow  P(\mathcal{A})=0$                                                                                                                                                                                                                                        &  \\[5ex] \cline{1-3}
Radon-Nikodyn derivative & $\frac{\mathrm{d}\widetilde\pi}{\mathrm{d}\pi}$,  see eq.~\ref{eq:RadonNikodym01} & $\frac{\mathrm{d}\widetilde P}{\mathrm{d}P}$, see eq.~\ref{eq:reweightPathSpace01}&  \\[5ex] \cline{1-3}
\end{tabular}
\end{table}

\subsection{Reweighting MSMs}
\label{sec:dynamicalReweighting}
When reweighting a MSM, we estimate the cross-correlation function $\widetilde C_{ij}(\tau)$ for the perturbed dynamics  (eq.~\ref{eq:perturbed_langevin}), 
from a set of paths of length $\tau$, $S_{\tau,m}=\lbrace \nu_1, \nu_2 ... \nu_m \rbrace$, which has been generated by the reference dynamics (eq.~\ref{eq:unperturbed}).
The cross-correlation function $\widetilde C_{ij}(\tau)$ is a combined phase space and path ensemble average (eq.~\ref{eq:pathEnsembleAverage02}).
Thus, both averages have to be reweighted with the appropriated probability ratio
\begin{eqnarray}
	\mathbb{E}_{\widetilde P, \widetilde \pi}[a(\mathbf{x}_0)f(\omega)]	
	&=& \int_{\Gamma}    a(\mathbf{x}_0)  \mu_{\widetilde \pi}(\mathbf{x}_0) \int_{\Omega_{\tau, \mathbf{x}}}
		f(\omega)  \,  \mu_{\widetilde P}(\omega) \, \mathrm{d}\omega \, \mathrm{d}\mathbf{x}_0 \cr
	&=& \int_{\Gamma}    a(\mathbf{x}_0) g(\mathbf{x}_0)  \mu_{\pi}(\mathbf{x}_0) \int_{\Omega_{\tau, \mathbf{x}}}
		f(\omega) \, M_{\tau, \mathbf{x}}(\omega) \,  \mu_{P}(\omega) \, \mathrm{d}\omega \, \mathrm{d}\mathbf{x}_0 \, .
\label{eq:crossCorrelationReweighted01}	
\end{eqnarray}
With $a(\mathbf{x}_0) = \mathbf{1}_{B_i}(\mathbf{x}_0)$ and $f(\omega)=\mathbf{1}_{B_j}(\mathbf{x}_n)$, we obtain
\begin{eqnarray}
	\widetilde C_{ij}(\tau) 
	&=&\mathbb{E}_{\widetilde P, \widetilde \pi}[\mathbf{1}_{B_i}(\mathbf{x}_0)\mathbf{1}_{B_j}(\mathbf{x}_n)]	ÊÊ\cr
	&=& \int_{\Gamma}    \mathbf{1}_{B_i}(\mathbf{x}_0) g(\mathbf{x}_0)  \mu_{\pi}(\mathbf{x}_0) \int_{\Omega_{\tau, \mathbf{x}}}
		\mathbf{1}_{B_j}(\mathbf{x}_n) \, M_{\tau, \mathbf{x}}(\omega) \,  \mu_{P}(\omega) \, \mathrm{d}\omega \, \mathrm{d}\mathbf{x}_0 \, .
\label{eq:crossCorrelationReweighted01}	
\end{eqnarray}
which can be estimated from a set of paths $S_{\tau,m}=\lbrace \nu_1, \nu_2 ... \nu_m \rbrace$ as
\begin{eqnarray}
	\widetilde C_{ij}(\tau)&=&
	\lim_{mÊ\rightarrow \infty }\frac{1}{m}\sum_{\nu_k \in  S_{\tau, m}}  
	    g([\mathbf{x}_0]_k) \mathbf{1}_{B_i}([\mathbf{x}_0]_k) \cdot M_{\mathbf{x}, \tau}(\nu_k) \mathbf{1}_{B_j}([\mathbf{x}_n]_k)	\,.
\label{eq:illustration08}	
\end{eqnarray}
where $[\mathbf{x}_i]_k$ is the $i$th time step of the $k$th path. 
As in eq.~\ref{eq:correlationMatrix}, the initial states of the paths are not fixed but are distrubuted according to the equilibrium distribution $\mu_{\pi}(\mathbf{x})$
of the unperturbed dynamics.
Finally, the transition probability between set $B_i$ and set $B_j$ for the perturbed dynamics is obtained as 
\begin{eqnarray}
	\widetilde T_{ij}(\tau) = \frac{\widetilde C_{ij}(\tau)}{\sum_{j} \widetilde C_{ij}(\tau)} \, . 
\label{eq:illustration09}	
\end{eqnarray}
The Radon-Nikodym derivative for path ensembles $g(\mathbf{x})$ contains the ratio of the partition functions $\nicefrac{Z}{\widetilde Z}$ as a multiplicative factor. 
Since this factor appears both in the numerator and the denominator of eq.~\ref{eq:illustration09}, it cancels, and the partition functions do not have to be calculated.
$T_{ij}(\tau)$ (and analogously $\widetilde T_{ij}(\tau)$) is an element of the $s \times s$ MSM transition matrix $\mathbf{T}(\tau)$, where $s$ is the number of disjoint sets (microstates). 
We characterize the MSM by plotting and analyzing the dominant left and right eigenvectors of the transitiom matrix
\begin{eqnarray}
	\mathbf{T}(\tau)\mathbf{r}_i		&=& \lambda_i(\tau) \mathbf{r}_i \cr
	\mathbf{l}_i^{\top} \mathbf{T}(\tau)	&=& \lambda_i(\tau)  \mathbf{l}_i^{\top} 
\label{eq:MSMEigenvectors}	
\end{eqnarray}
where $\mathbf{l}_i^{\top}$ denotes the transpose of vector $\mathbf{l}_i$. 
We assess the approximation quality of the MSM by checking wether the implied timescales
\begin{eqnarray}
	t_i	&=& -\frac{\tau}{\ln(\lambda_i(\tau))}	= \mbox{const} \quad \forall \,  \tau > 0
\label{eq:ITS}	
\end{eqnarray}
are constant \cite{Swope2004, Prinz2011}.

\subsection{Projection}
\label{sec:projection}
We now consider a perturbation $U(\cdot)$ that does not directly affect the relevant coordinates used to construct the MSM. 
In such a situation the perturbation acts mainly on the coordinates directly perturbed and has a minor effect on the other degrees of freedom, in particular on the relevant coordinates that do not capture the full effect of the perturbation.
Thus the reweighting may become problematic because the reweighting 
formula \eqref{eq:reweightPathSpace01} is dominated by large, fluctuating gradients, therefore a rescaling of the force needs to be introduced.
To address this issue, we propose to project the gradient of the perturbation onto the coordinates used to construct the MSM. Let's assume that the MSM has been built on a combination of $d$ coordinates $\chi_1,...\chi_d$, then the equation \ref{eq:reweightPathSpace01} is rewritten as:
\begin{equation}
M_{\tau, \mathbf{x}}(\omega)
= \frac{\mu_{\widetilde{P}}(\omega)}{\mu_P (\omega)} = \exp \left\lbrace - \sum_{i=1}^{3N} \left[ \sum_{k=0}^n \frac{\mathbf{c}_{i,k}}{\sigma_i} \eta^i_k \sqrt{\Delta t} - \frac{1}{2} \sum_{k=0}^n  \left(\frac{\mathbf{c}_{i,k}}{\sigma_i} \right)^2 \Delta t\right] \right\rbrace
\label{eq:reweighting_projection}
\end{equation}
with 
\begin{equation}
\mathbf{c}_{i,k} =  \sum_{j=1}^d \frac{\langle\nabla_i U(\mathbf{r}_k) , \chi_{j,k} \rangle}{\langle \chi_{j,k} , \chi_{j,k} \rangle} \chi_{j,k} 
 \end{equation}

\section{Methods}
\subsection{Two dimensional system}
The Brownian dynamics on a two-dimensional potential energy function $V(x,y)$ 
\begin{equation}
\begin{cases}
\mathrm{d} x_t=-\nabla_x V\left(x_t, y_t\right) + \sigma \mathrm{d} B_t^x \\
\mathrm{d} y_t=-\nabla_y V\left(x_t, y_t\right) + \sigma \mathrm{d} B_t^y
\label{eq:BrownianDynamics}
\end{cases}
\end{equation}
have been solved using the Euler-Maruyama scheme\citep{b9} with an integration time step of $\Delta t=0.001$. 
The term $B_t^i$ denotes a standard Brownian motion in direction $i=x,y$, $\sigma=1$ is the volatility, and 
the random variables $\eta^i$ were drawn from a standard Gaussian distribution. 
The reference potential energy function was
\begin{equation}
V(x,y) = (x^2-1)^2+(y^2-1)^2 + |x-y| \, ,
\label{eq:2D_referencePotential}
\end{equation}
and the perturbed potential energy function was $\widetilde V(x,y,)  = V(x,y) + U(y)$ with
\begin{eqnarray}
	U(y) = - y	
\label{eq:2D_perturbedPotential}
\end{eqnarray}
For both potential energy functions, trajectories of $8 \times 10^7$ time-steps were produced. 
In both simulations, the path probability ratio $M_{\tau, \mathbf{x}}$ was calculated using eq.~\ref{eq:reweightPathSpace01}.
The MSM have been constructed by discretizing each dimension $x$ and $y$ into 40 bins, yielding 1600 microstates. The chosen lagtime was $\tau=400$ time-steps.

%
%
\subsection{Many-body system in three-dimensional space}
We designed a six-particle system, in which five particles form a chain while the sixth particle branches the chain at the central atom 
(Fig.~\ref{graph:manybody}.A).
The position of the $i$th particle is $\mathbf{r}_i \in \mathbb{R}^3$.
The potential energy between two directly bonded atoms (blue lines in Fig.~\ref{graph:manybody}.A) was
\begin{equation}
V \left(\mathbf{r}_{ij} \right) = \left( \mathbf{r}_{ij}^2 -1 \right)^2 + 0.6\mathbf{r}_{ij}
\end{equation}
with $\mathbf{r}_{ij} = \mathbf{r}_j - \mathbf{r}_i$.
The bond potential energy function is a tilted double well potential. 
This ensures that the potential energy function of the complete system has multiple minima with varying depths. 
No non-bonded interactions were applied. Thus, the reference potential energy function of the complete system was
\begin{equation}
V \left(\mathbf{r} \right) = \sum_{ij=12,13,14,24,36 }V \left(\mathbf{r}_{ij} \right) 
\end{equation}
The Langevin dynamics (eq.~\ref{eq:unperturbed}) of this system have been solved using the BBK integrator \citep{b5} with an integration time step of $\Delta t=0.001$.
The masses $M$ of the particles, temperature $T$, the friction coefficient $\gamma$, and the Boltzmann constant $k_B$ were all set to one.
The perturbed potential energy function was $\widetilde V(\mathbf{r})  = V(\mathbf{r}) + U(\mathbf{r})$ with
\begin{equation}
U \left(\mathbf{r} \right) = \frac{1}{2} \mathbf{r}_{24}^2 + \frac{1}{2} \mathbf{r}_{34}^2
\end{equation}
The perturbation is a harmonic potential energy along the through-space distance between atoms $(2,4)$ and $(3,4)$, respectively 
(green dashed line in Fig.~\ref{graph:manybody}.A).
For both potential energy functions, trajectories of $3.2 \times 10^8$ time-steps were produced. 
The MSMs were constructed on the two-dimensional space spanned by the bond-vectors $\mathbf{r}_{12}$ and $\mathbf{r}_{13}$, i.e. on two coordinates which were not directly perturbed. Each dimension $x$ and $y$ has been discretized into 40 bins, yielding 1600 microstates. The chosen lagtime was $\tau=400$ time-steps.
In both simulations, the path probability ratio $M_{\tau, \mathbf{x}}$ was calculated by projecting the gradient vectors $\nabla U \left(\mathbf{r}_{24} \right)$ and $\nabla U \left(\mathbf{r}_{34} \right)$ on the vectors $\mathbf{r}_{12}$ and $\mathbf{r}_{13}$ and subsequently evaluating 
eq.~\ref{eq:reweighting_projection}.

%
%
\subsection{Alanine and Valine dipeptide}
We performed all-atom MD simulations of acetyl-alanine-methylamide (Ac-A-NHMe, alanine dipeptide) in implicit and explicit water and of acetyl-valine-methylamide (Ac-V-NHMe, valine dipeptide) in implicit water. All simulations were carried out with the OPENMM 7.01 simulation package\citep{b7}, in an NVT ensemble at 300 K. Each system was simulated with the force field AMBER ff-14sb \citep{b6}. The water model was chosen according to the simulation, i.e. the GBSA model \citep{Onufriev2004} for implicit solvent simulation and the TIP3P model \citep{b8} for explicit solvent simulation. For each of these setups, the aggregated simulation time was 1 $\mu$s and we printed out the positions every \textsf{nstxout}=100 time steps, corresponding to 0.2 ps. A Langevin thermostat has been applied to control the temperature and a Langevin leapfrog integrator\cite{Izaguirre2010} has been used to integrate eq.~\ref{eq:unperturbed}. For implicit solvent simulations, interactions beyond 1 nm are truncated. For explicit solvent simulations, periodic boundary conditions are used with the Particle-Mesh Ewald (PME) algorithm to estimate Coulomb interactions.

In the alanine dipeptide simulations, we have perturbed the potential energy function of the backbone dihedral angles 
$\phi$ and $\psi$.
The reference potential energy functions were
\begin{equation}
V(\phi) = 0.27\cos(2\phi) + 0.42 \cos(3\phi) 
\label{eq:peturbation_phi}
\end{equation}
\begin{equation}
V(\psi) = 0.45\cos(\psi-\pi) + 1.58 \cos(2\psi-\pi) + 0.44\cos(3\psi-\pi) 
\label{eq:peturbation_psi}
\end{equation}
where the parameters have been extracted from force field files and $\pi$ denotes the mathematical constant. 
The perturbed potential energy function was a harmonic potential along each dihedral angle degree of freedom
\begin{equation}
\label{eq:perturbation_alanine}
U(\phi, \psi) =  \frac{1}{2} \kappa_\phi \phi^2 + \frac{1}{2} \kappa_\psi \psi^2
\end{equation}
where $\kappa_\phi$ and $\kappa_\psi$ are the force constants, which could be adjusted after the simulation 
(Fig.~\ref{graph:alanine}.A and \ref{graph:alanine}.B).
The gradient $\nabla_{\mathbf{r}_i} U(\cdot)$ in eq.~\ref{eq:reweightPathSpace01} is defined with respect to the
Cartesian coordinates. 
Thus, applying the chain rule, the path probability ratio for the perturbation of the backbone dihedral angles is given as
\begin{eqnarray}
\label{eq:girsanov_k}
M_\tau &=& \exp \lbrace - \sum_i^{N} [
			\kappa_\phi \int_0^\tau \frac{\phi(s)}{\sigma_i} \frac{\partial \phi(s)}{\partial \mathbf{r}_i}\mathrm{d}B_s^i +
			\kappa_\psi \int_0^\tau \frac{\psi(s)}{\sigma_i} \frac{\partial \psi(s)}{\partial \mathbf{r}_i}\mathrm{d}B_s^i - \cr\cr
	&&		\frac{1}{2} \kappa_\phi^2 \int_0^\tau \left(\frac{\phi(s)}{\sigma_i} \frac{\partial \phi(s)}{\partial \mathbf{r}_i} \right)^2 \mathrm{d}s -
			\frac{1}{2} \kappa_\psi^2 \int_0^\tau \left(\frac{\psi(s)}{\sigma_i} \frac{\partial \psi(s)}{\partial \mathbf{r}_i} \right)^2 \mathrm{d}s \cr\cr
	&& 		\kappa_\phi \kappa_\psi \int_0^\tau \frac{\phi(s) \psi(s)}{\sigma_i^2} \frac{\partial \phi(s)}{\partial \mathbf{r}_i} \frac{\partial \psi(s)}{\partial \mathbf{r}_i}\mathrm{d}s 
] \rbrace \, .
\end{eqnarray}

In the valine dipeptide simulation, we have perturbed the $\chi_1$ side-chain dihedral angle.
The reference potential energy function was
\begin{equation}
V(\chi) = 0.337\cos(\chi) + 0.216\cos(2\chi-\pi) + 0.001\cos(4\chi-\pi) + 0.148\cos(3\chi) \,.
\end{equation}
where the parameters have been extracted from force field files and $\pi$ denotes the mathematical constant. 
The perturbed potential energy function was a harmonic potential 
\begin{equation}
\label{eq:perturbation_valine}
U(\chi) =  \frac{1}{2} \kappa_\chi \chi^2 
\end{equation}
where $\kappa_\chi$ is the force constant, which could be adjusted after the simulation (Fig.~\ref{graph:valine}.A).
Thus, the path probability ratio is given as
\begin{eqnarray}
\label{eq:girsanov_k}
M_\tau &=& \exp \lbrace - \sum_i^{N} [
			\kappa_\chi \int_0^\tau \frac{\chi(s)}{\sigma_i} \frac{\partial \chi(s)}{\partial \mathbf{r}_i}\mathrm{d}B_s^i -
			\frac{1}{2} \kappa_\chi^2 \int_0^\tau \left(\frac{\chi(s)}{\sigma_i} \frac{\partial \chi(s)}{\partial \mathbf{r}_i} \right)^2 \mathrm{d}s 
] \rbrace \, .
\end{eqnarray}
The MSM, for both alanine dipeptide and valine dipeptide simulations, have been constructed  by discretizing the dihedral angles $\phi$ and $\psi$ into 36 bins each, yielding  1296 microstates. The chosen lagtime was 20 ps for both the implicit and explicit solvent simulation.

\section{Results and discussion}

\subsection{Efficient implementation}
\label{sec:implementation}
To estimate a MSM at a perturbed potential energy function $V(\mathbf{x}) + U(\mathbf{x}, \kappa)$ with the dynamical reweighting method, one simulates a long trajectory $\mathbf{x}(t)$ at a reference potential energy function $V(\mathbf{x})$
using an integration time step $\Delta t$.
From this trajectory $m$ short paths of length $\tau  = n \Delta t$ are extracted, yielding a set of paths $S_{\tau, m} = \lbrace \nu_1, \nu_2, ... \nu_m\rbrace$, 
which can subsequently be used to evaluate eq.~\ref{eq:illustration08}, 
where $g([\mathbf{x}_0]_k)$ is given by eq.~\ref{eq:RadonNikodym01} and $M_{\mathbf{x},\tau}(\nu_k)$ is given by eq.~\ref{eq:reweightPathSpace01} or eq.~\ref{eq:reweighting_projection}.
As discussed in section \ref{sec:dynamicalReweighting}, the factor $\nicefrac{Z}{\widetilde Z}$ cancels for the estimate of $\widetilde T_{ij}(\tau)$ (eq.~\ref{eq:illustration09})
and the partition functions do not need to be calculated.
Let us assume that the simulation integrates the Langevin equation of motion for the reference potential energy function $V(\mathbf{x})$ (eq.~\ref{eq:unperturbed}).
To estimate a MSM with transition probabilities $\widetilde T_{ij}(\tau)$ for the dynamics in the perturbed potential energy function from a set of paths
$S_{\tau, m} = \lbrace \nu_1, \nu_2, ... \nu_m\rbrace$, we need to know
the value of the perturbation $U([\mathbf{x}_t]_k, \kappa)$ at every time step $t$  and for each path $\nu_k$, 
the gradient of the perturbation $\nabla U([\mathbf{x}_t]_k, \kappa)$ at every time step $t$  and for each path $\nu_k$, 
the random numbers $\eta_k$ generated at every time step of the simulation of each path $\nu_k$, and 
the volatility $\sigma$. 
The volatility is determined by the temperature and the friction coefficient (eq.~\ref{eq:EinsteinRelation}), both of which are input parameters for the simulation algorithm.
To calculate the other three properties one needs to know the positions and the random numbers at every integration time step of the simulation. 
In a naive implementation of the reweighting method one would hence write out the positions and random number at every time step and calculate $g([\mathbf{x}_0]_k)$ and $M_{\mathbf{x},\tau}(\nu_k)$
in a post-analysis step.
The advantage of this approach is that the set of paths can be reweighted to the path probability measure of any perturbation $U(\mathbf{x}, \kappa)$, 
as long as the absolute continuity is respected (eq.~\ref{eq:absoluteContinuity01}).
On the other hand, this approach is hardly practical because writing out a trajectory at every integration time step quickly fills up any hard disc
and slows down the simulation considerably.
We therefore decided to compute the probability ratios $g([\mathbf{x}_0]_k)$ and $M_{\mathbf{x},\tau}(\nu_k)$  ``on the fly'' during the simulation. 
In practice, the lag time $\tau$ of a MSM can only assume values integer multiples of the frequency \textsf{nstxout} at which the positions are written to file,
i.e. of $\tau = n \Delta t = A \cdot \mbox{\textsf{nstxout}} \cdot \Delta t$ with $A \in \mathbb{N}$.
The discretized It$\hat{\mathrm{o}}$ integral and the discretized Riemann integral in eq.~\ref{eq:reweightPathSpace01} 
are sums from time step $k=0$ to $k=n$, which can be broken down into $A$ sums of size $\textsf{nstxout}$
\begin{eqnarray}
	\sum_{k=0}^n \dots &=& 
	\sum_{k=0}^{\mbox{\textsf{\scriptsize nstxout}}-1} \dots + 
	\sum_{k=\mbox{\textsf{\scriptsize nstxout}}}^{2\cdot \mbox{ \textsf{\scriptsize nstxout}}-1} \dots +
	\sum_{k=(A-1)\cdot \mbox{\textsf{\scriptsize nstxout}}}^{A\cdot \mbox{\textsf{\scriptsize nstxout}}-1}	
\end{eqnarray}
Thus, we calculate the terms 
\begin{eqnarray}
	I(a) = - \sum_{i=1}^{3N} \sum_{k=(a-1)\cdot \mbox{\textsf{\scriptsize nstxout}}}^{a\cdot \mbox{\textsf{\scriptsize nstxout}}-1}\frac{\nabla_i U(\mathbf{r}_k)}{\sigma} \eta^i_k \sqrt{\Delta t}
\end{eqnarray}
and 
\begin{eqnarray}
	R(a) = - \sum_{i=1}^{3N} \frac{1}{2} \sum_{k=(a-1)\cdot \mbox{\textsf{\scriptsize nstxout}}}^{a\cdot \mbox{\textsf{\scriptsize nstxout}}-1} \left(\frac{\nabla_i U(\mathbf{r}_k)}{\sigma} \right)^2 \Delta t
\end{eqnarray}
``on the fly''  and write out the results at the same frequency \textsf{nstxout} as the positions. 
The path probability ratio is reconstructed after the simulation as
\begin{eqnarray}
	M_{\tau, \mathbf{x}}(\omega) &=& \exp \left\lbrace \sum_{a=1}^A I(a) + R(a) \right\rbrace \, .
\end{eqnarray}
The potential energy of the perturbation $U(\mathbf{x}_t)$ is written out at the frequency \textsf{nstxout}  and the complete weight $g([\mathbf{x}_0]_k) \cdot M_{\mathbf{x},\tau}(\nu_k)$ is calculated during the construction of the MSM.
The lag-time $\tau$ can be chosen and varied after the simulation. 
The modification of the MD integrator can be readily implemented within the MD software package OpenMM \citep{b7}. An example script is provided in the supplementary material. 

The approach requires that the perturbation potential energy function $U(\mathbf{r}, \kappa)$ is chosen prior to the simulation. 
Note however that, 
if the perturbation potential energy function depends linearly on the parameter $\kappa$,  
i.e. if $\kappa$ is a force constant $U(\mathbf{r}, \kappa) = \kappa\cdot U(\mathbf{r}) $, 
then so do the two integrals 
$I(a, U(\mathbf{r}, \kappa)) = \kappa\cdot I(a, U(\mathbf{r}))$ and
$R(a, U(\mathbf{r}, \kappa)) = \kappa\cdot R(a, U(\mathbf{r}))$.
Thus, it is sufficient to calculate $I(a, U(\mathbf{r}))$ and $R(a, U(\mathbf{r}))$ ``on the fly'' and to scale the integrals after the simulation to any desired value of $\kappa$.
A single simulation is sufficient to allow for reweighting to a whole series of perturbation potential energy functions. 
Also, the integrator can be modified such that the It$\hat{\mathrm{o}}$ integrals $I_1(a), I_2(2)... $ and the Riemann integral $R_1(a), R_2(2)... $
of several functionally different perturbation energy functions $U_1(\mathbf{x}), U_2(\mathbf{x})... $ are calculated.
Thus, using a single reference simulation one can reweight to several  functionally different perturbations and scale these perturbations by an arbitrary force constant.

\begin{figure}[h!]
  \centering
  \includegraphics[scale=1]{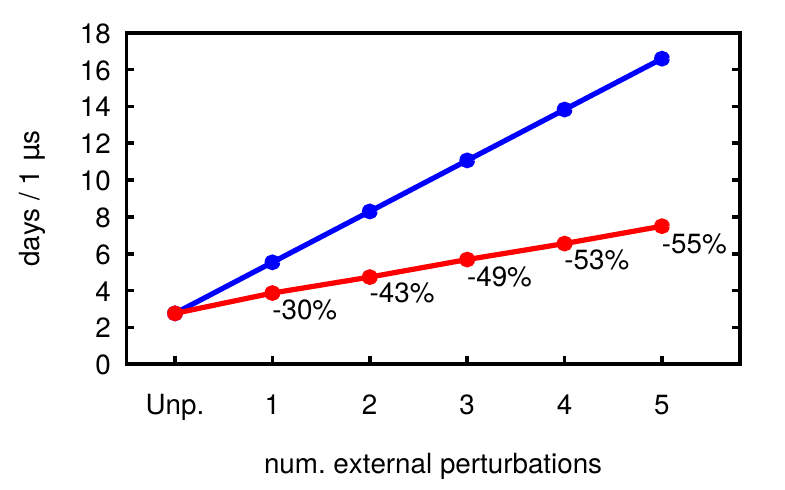}
    \caption{Number of days needed to product a trajectory of Alanine dipeptide in implicit solvent of 1 $\mu$s. The system has been perturbed by adding an harmonic potential to different dihedral angles. The blue line denotes the time needed to perform respectively 1,...,5 simulations with different potential energy functions (i.e. different perturbations). The red line is the time necessary to perform one single simulation and to compute the Girsanov formula on fly for 1,...,5 different perturbations at the same time. The benchmark test has been realized on a CPU \textsf{Intel(R) Core(TM) i5-4590 CPU @ 3.30GHz} with 15 GB of RAM. 
}
        \label{graph:benchmark}
\end{figure}

The computional cost of adding the calculation of $I_i(a)$ and $R_i(a)$ for a perturbation energy function $U_i(\mathbf{x})$ is modest.
The blue line in Fig.~\ref{graph:benchmark} shows the computational costs for simulating alanine dipeptide in implicit water with $n=0,1,2,3,4,5$ different perturbations as the number of days required to obtain a trajectory of 1 $\mu$s for each perturbation on a small workstation.
The red line in Fig.~\ref{graph:benchmark} shows the computational cost of implementing the same $n$ perturbations into a single reference simulation.
For a single perturbation 30\% of the computational cost is saved by dynamical reweighting, whereas for 5 perturbations more than half of the computational cost is saved.
The gain is even greater, if the dependence on a force constant is to be studied. 
Moreover, in simulation boxes of larger systems with explicit solvent, the cost of evaluating of the perturbation potential energy is small compared to 
the normal force evaluations, such that for these systems the computional cost of calculating the $I_i(a)$ and $R_i(a)$ ``on the fly'' becomes negligible.
We remark that if $I_i(a)$ and $R_i(a)$ are too large, the Girsanov reweighting method might become numerically intractable.

\subsection{Stochastic forces}
\label{sec:stochasticForces}
The dynamical reweighting method has been derived by using the Langevin equation of motion as starting point. 
The concept of the path probability is inextricably linked to the presence of Gaussian random forces in the equation of motion (see appendix \ref{sec:continuousPaths}).
However from a physical perspective, the Langevin dynamics only approximates the Hamiltonian dynamics of the complete system, 
by splitting the system in a subsystem $S$ and a heat bath $B$ and replacing the interaction of the subsystem with the heat bath by a friction term and a stochastic process.
In most MD simulations, thermostatted versions of the Hamiltonian dynamics are propagated in which no stochastic forces are generated.
In principle, one could split such a simulation box into two subsystem $S$ and $B$ and use the forces which subsystem $B$ exerts on subsystem $S$ as 
substitute for the random forces.
Unfortunately, even if the heat bath $B$ consists of Lennard-Jones particles, the forces on $S$ deviate too much from a Gaussian random process to be used 
in the dynamical reweighting method (data not shown). 
We therefore decided to use the Langevin leapfrog integrator\cite{Izaguirre2010} to integrate the Langevin equation of motion for 
both the implicit and explicit solvent simulations and used the random forces generated by the integrator to reweight the path ensemble. 
\subsection{Two dimensional system}
\label{sec:twodimsys}
\begin{figure*}[h!]
  \begin{center}
  \includegraphics[width=16cm]{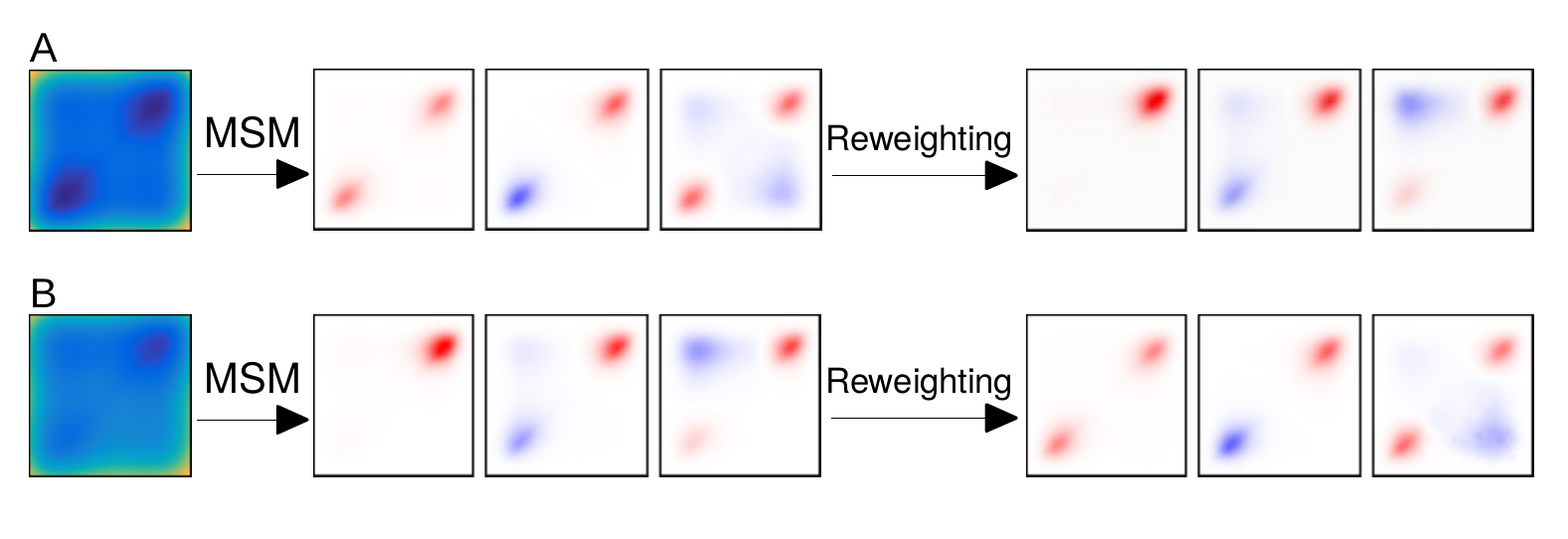}
    \caption{(A) Reference potential. (B) Dominant left eigenvectors of the reference potential. (C) Dominant left eigenvectors reweighted from the reference dynamics. (D) Perturbed potential. (E) Dominant left eigenvectors of the perturbed potential. (F) Dominant left eigenvectors reweighted from the perturbed dynamics.}
    \label{graph:eigenvectors_2d}
  \end{center}
\end{figure*}

As a first application, we consider the Brownian dynamics of a particle moving on a two-dimensional potential energy function (eq.~\ref{eq:BrownianDynamics}). 
The reference potential energy function $V(x,y)$ (eq.~\ref{eq:2D_referencePotential}, Fig.~\ref{graph:eigenvectors_2d}.A) has two minima at $(-1,-1)$ and $(1,1)$ which are connected by a transition state at $(0,0)$.
We added at a perturbation (eq.~\ref{eq:2D_perturbedPotential}) which tilts the energy function $\widetilde V(x,y)$ along the direction $y$, such that 
the mininum at $(1,1)$ becomes much deeper than the minimum at $(-1,-1)$.
Fig.~\ref{graph:eigenvectors_2d}.B and \ref{graph:eigenvectors_2d}.E show the dominant left MSM eigenvectors of the two systems (direct MSMs).
The first eigenvector corresponds to the equilibrium distribution. In both cases, the second eigenvector represents the transition between the two wells, while the third eigenvector corresponds to an exchange of probability density between the transition state region and the two wells.
Fig.~\ref{graph:eigenvectors_2d}.C shows the MSM eigenvectors obtained by reweighting the simulation in $V(x,y)$  
to the perturbed potential energy function $V(x,y)+U(y)$ (reweighted MSM). 
Both eigenvectors (eq.~\ref{eq:MSMEigenvectors}) and implied timescales (eq.~\ref{eq:ITS}) are in perfect agreement with Fig.~\ref{graph:eigenvectors_2d}.E.
This confirms that reweighting MSMs using the Girsanov formula works well for low-dimensional Brownian dynamics\cite{Schuette2015b}.  
Note that the simulation in the reference potential energy function $V(x,y)$ exhibits frequent transitions between the two minima, and thus eq.~\ref{eq:absContPath} is certainly fulfilled. 
By contrast in the perturbed system $\widetilde V(x,y)$, the simulation sampled considerably fewer transitions between the minima and only a small fraction of the simulation time was spent in the minimum in the lower left corner.
Fig.~\ref{graph:eigenvectors_2d}.F shows the dominant eigenvectors of a MSM of the reference dynamics constructed by reweighting the perturbed path ensemble. 
The fact that Fig.~\ref{graph:eigenvectors_2d}.C is in excellent agreement with Fig.~\ref{graph:eigenvectors_2d}.B demonstrates that the reweighting method yields accurate 
results even for path ensembes with low numbers of transitions across the largest barriers of the system and thus suggests that it can be applied to high-dimensional dynamics on rugged potential energy surfaces.

%
%
\subsection{Many-body system in three-dimensional space}
\label{sec:manybody}
\begin{figure*}[h!]
  \begin{center}
  \includegraphics[scale=1]{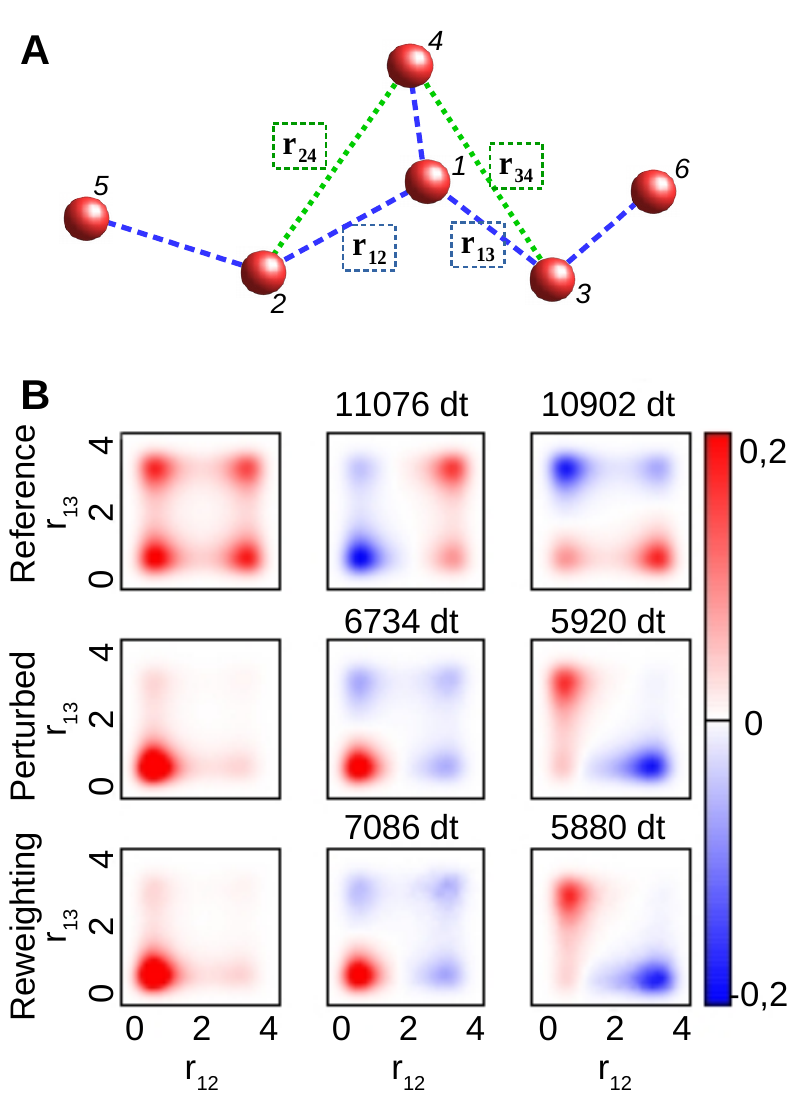}
    \caption{Many body system. (A) The system is perturbed adding the bonds $\mathbf{r_{24}}$ and $\mathbf{r_{34}}$, while the MSM is constructed on the distances $\mathbf{r_{12}}$ and $\mathbf{r_{13}}$. The numbers written in italic identify the atoms of the molecule. (B) Dominant eigenvectors of the manybody system. (first row) Unperturbed system; (second row) perturbed system (third row) reweighting method. The number above the eigenvectors denote respectively the implied time scales of the second and third process as number of timesteps.}
    \label{graph:manybody}
  \end{center}
\end{figure*}

As a second example, we studied a six-particle system, in which five particles form a chain while the sixth particle branches the chain at the central atom Fig.~\ref{graph:manybody}.A .
We constructed the MSM on the central bonds $\mathbf{r}_{12}$ and $\mathbf{r}_{13}$, but applied the perturbation potential energy function along the through-space distances
$\mathbf{r}_{24}$ and $\mathbf{r}_{34}$.  
Thus, the perturbation was projected onto the reaction coordinates $\mathbf{r}_{12}$ and $\mathbf{r}_{13}$ during the reweighting.
The reference potential energy function for each bond is a tilted double well potential, such that the reference system exhibits four metastable states
(first row of Fig.~\ref{graph:manybody}.B).
The six-particle system and the reference potential energy function is symmetric. 
Thus the MSM of the reference system has two degenerate dominant eigenfunction with implied timescales of $1.1\cdot 10^4 \Delta t$. 
with a relative error due to sampling of 1.6 \%.
The perturbation contracts the bonds, thereby stabilizing the metastable state at the lower left corner in the 1st eigenvector (second row of Fig.~\ref{graph:manybody}.B).
Its effect  is to break the symmetry and to accelerate the dynamics, yielding implied timescales of $6.7\cdot 10^4 \Delta t$ and $5.9 \cdot 10^4 \Delta t$.
The third row of Fig.~\ref{graph:manybody}.B shows the eigenvectors and the implied time scales obtained by reweighting the simulation at the reference potential energy function to the perturbed potential energy function. 
The relative error of the implied timescale of the second and third eigenvector is 5.2\% and 0.7\%, respectively.
%

%
%
\subsection{Alanine dipeptide and valine dipeptide}
\label{sec:alanineValine}
\begin{figure*}[h!]
    \begin{center}
  \includegraphics[scale=1]{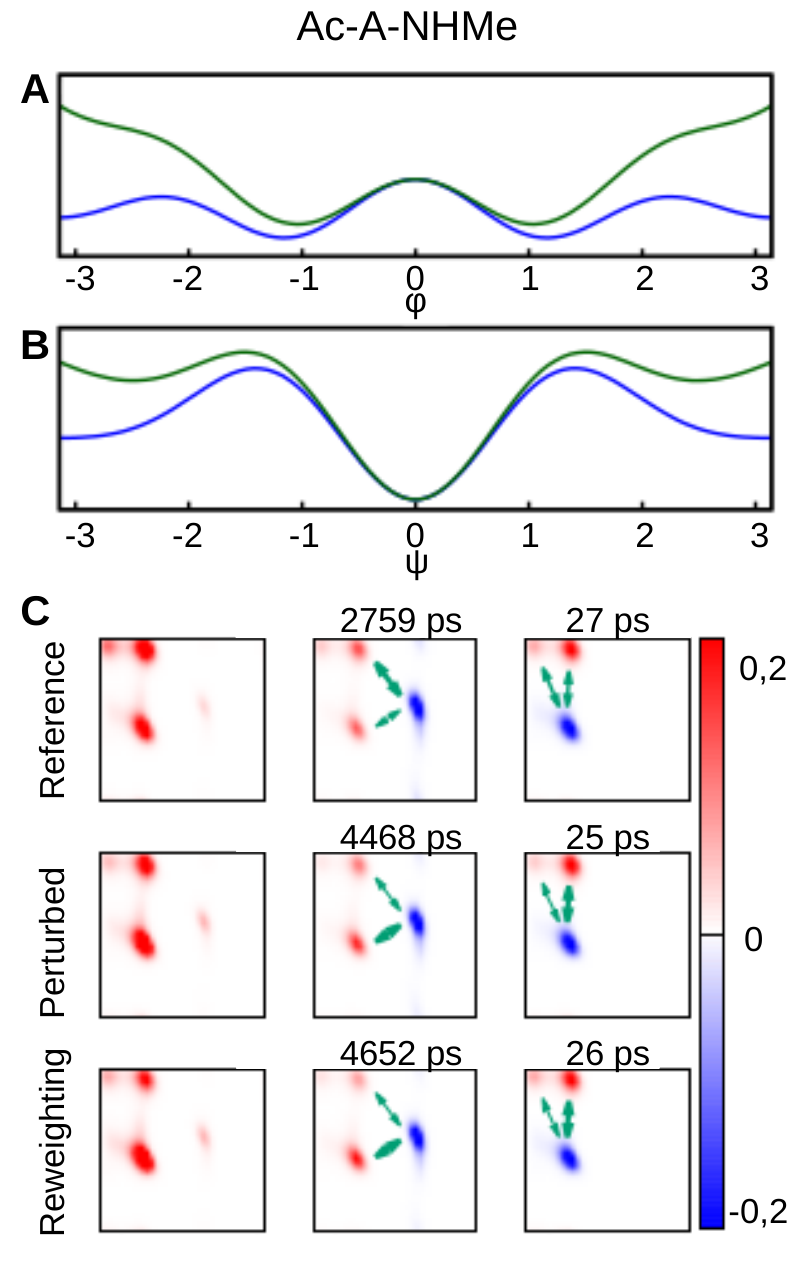}
    \caption{Alanine dipeptide.
(A) 	Reference potential energy function of the backbone dihedral angle $\phi$,
	$V(\phi)$ (blue line), and perturbed potential energy function, $V(\phi) + U(\kappa_\phi=0.5, \kappa_\psi=0, \phi, \psi)$ (green line). 
(B) Reference potential energy function of the backbone dihedral angle $\psi$,
	$V(\psi)$ (blue line), and perturbed potential energy function,  $V(\psi) + U(\kappa_\phi=0, \kappa_\psi=0.5, \phi, \psi)$ (green line).    
(C) First three MSM eigenvectors of alanine dipeptide, where the MSM is constructed in the space spanned by the $\phi$- and $\psi$-backbone dihedral angle.
\label{graph:alanine}}
    \end{center}
\end{figure*}

Figure ~\ref{graph:alanine} shows the results for alanine dipeptide (Ac-A-NHMe) in implicit water.
The MSM has been constructed on the $\phi$ and $\psi$ backbone dihedral angles and the slow eigenvectors of the unperturbed system are shown in 
Fig.~\ref{graph:alanine}.C.
The first eigenvector shows the typical equilibrium distribution in the Ramachandran plane \cite{Vitalini2015}. 
The second eigenvector represents a torsion around the $\phi$ angle and corresponds to a kinetic exchange between the L$_{\alpha}$-minimum ($\phi>0$)
and the $\alpha$-helix and $\beta$-sheet minima ($\phi < 0$).
The associated timescale is 2.8 ns.
The green arrows in Fig.~\ref{graph:alanine}.C represent the frequency of the transitions, with the transition L$_{\alpha} \leftrightarrow \beta$-sheet conformation occuring more frequently than the transition 
 L$_{\alpha} \leftrightarrow \alpha$-helical conformation.
The third eigenvector represents a transition $\beta$-sheet $ \longleftrightarrow \alpha$-helical conformation, i.e. a torsion around $\psi$, 
and is associated to a timescale of 27 ps.
We perturbed the dynamics by adding a harmonic potential to the dihedral angle potentials of the $\phi$- and $\psi$-angle (Fig.~\ref{graph:alanine}.A and \ref{graph:alanine}.B, 
eq.~\ref{eq:perturbation_alanine} with $\kappa_\phi=0.5$ and $\kappa_\psi=0.5$).
The $\alpha$-helical region is somewhat stabilized by the perturbation but otherwise the dominant eigenvectors are very similar to the unperturbed system
(second row in Fig.~\ref{graph:alanine}.C). 
However, the perturbation changes the relative frequency of the two possible transitions (green arrows) in the second eigenvector, 
resulting in an increased implied timescale of 4.5 ns.
The third row of Fig.~\ref{graph:alanine}.C shows the dominant eigenvector of the perturbed system obtained by reweighting the reference simulations.
The results are in excellent agreement with the direct simulation of the perturbed systems. The relative error of the implied timescale associated to the second eigenvector is
4.1\%.
%

\begin{figure*}[h!]
    \begin{center}
  \includegraphics[scale=1]{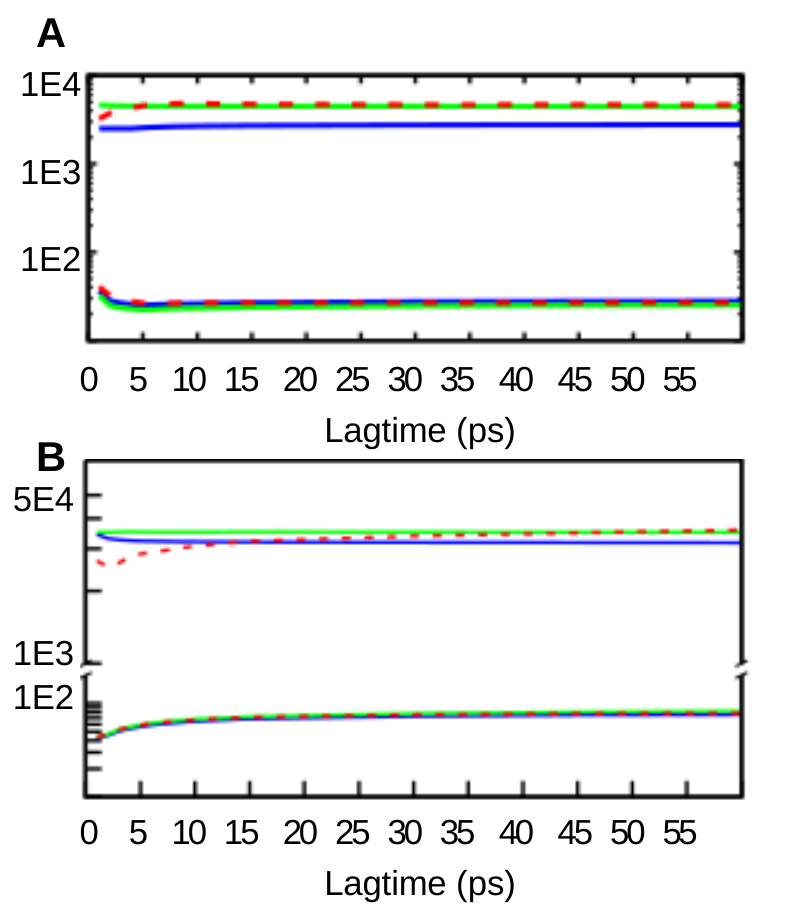}
    \caption{Alanine dipeptide.
(A) Implied time scales in implicit solvent. (B) Implied time scales in explicit solvent. Reference potential (blue line), perturbed potential (green line) and reweighting method (red dashed line).
   \label{graph:alanine_its}
}
    \end{center}
\end{figure*}

Fig.~\ref{graph:alanine_its}.A shows the implied timescale test (eq.~\ref{eq:ITS}) for alanine dipeptide in implicit water. 
All three systems (reference, perturbed and reweighted) show constant implied timescales, indicating that the MSM are well converged.
Moreover, the graph shows that the reweighting method can recover the implied timescales of the perturbed system over a large range of lag times $\tau$.
We have repeated the alanine dipeptide simulations in explicit water. 
The eigenvectors are very similar to those of alanine dipeptide in implicit water (data not shown), but the associated implied timescales differ from the implicit solvent simulations
(Fig.~\ref{graph:alanine_its}.B). 
In the unperturbed system, the implied timescale of the second eigenvector was 3.1 ns and the implied timescale of the third eigenvector was 75 ps.
Perturbing the dihedral angle potentials, slightly increased the implied timescale of the second eigenvector to 3.5 ns and left the implied timescale of the third
eigenvector unaffected. 
The implied timescales of the reference simulation and the direct simulation of the perturbed system were constant, whereas we noticed a slight drift in the implied timescale of the second eigenvector for the reweighted MSM. 
At $\tau=45$ ps, the measured implied timescale is 3.4 ns which corresponds to a relative error of 2.8 \%.

\begin{figure*}[h!]
    \begin{center}
  \includegraphics[scale=1]{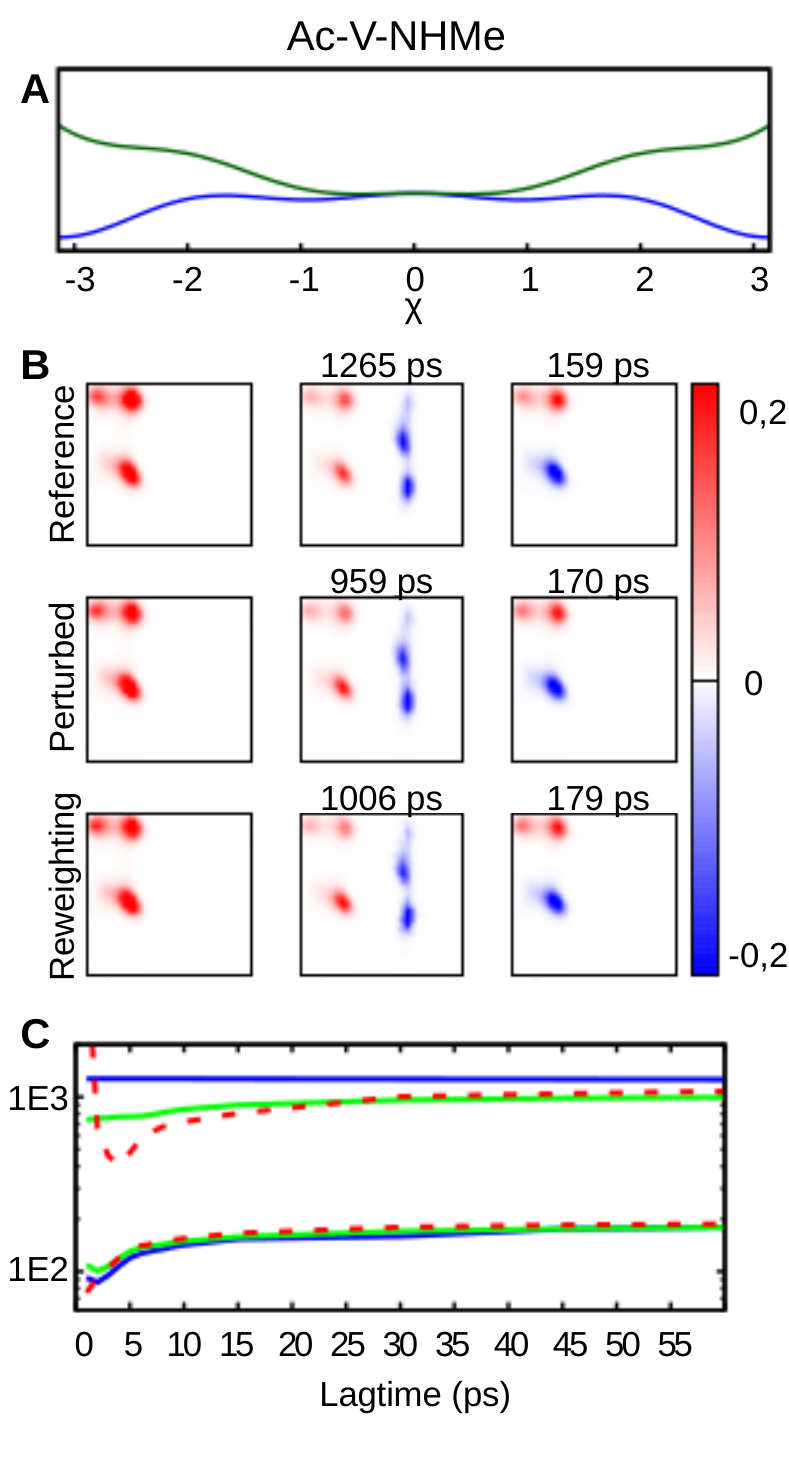}
    \caption{Valine dipeptide.
(A) Reference potential energy function of the side-chain dihedral angle $\chi$, $V(\chi)$ (blue line), and perturbed potential energy function, $V(\chi) + U(\kappa_\chi=0.5,\chi)$ (green line). 
(B) First three MSM eigenvectors of the valine dipeptide, where the MSM is constructed in the space spanned by the $\phi$- and $\psi$-backbone dihedral angle. 
(C) Associated implied time scales as a function of the lag time $\tau$. Reference potential (blue line), perturbed potential (green line) and reweighting method (red dashed line)
        \label{graph:valine}
}
    \end{center}
\end{figure*}

We also tested the dynamical reweighting method on valine dipeptide (Fig.~\ref{graph:valine}).
Here, however, we perturbed the potential energy function of the $\chi$-side chain dihedral angle (Fig.~\ref{graph:valine}.A) by adding a harmonic potential, 
while constructing the MSM on the $\phi$ and $\psi$ backbone dihedral angles. Thus, the perturbation did not directly act on the variables of the MSM.  
The perturbation of the $\chi$ angle had no affect on the eigenvectors of the MSM (Fig.~\ref{graph:valine}.B), but it did change the implied time scales. 
It caused a decrease of the implied time scale of the second eigenvector from 1.3 ns in the reference simulation to 1.0 ns in the perturbed simulation
and a slight increase of the implied time scale of the third eigenvector from 159 ps in the reference simulation to 170 ps in the perturbed simulation. 
Reweighting the reference simulation to the perturbed potential energy function recovered the results of the direct simulation of the perturbed system.
The implied timescales obtained by the reweighting calculation were 1.0 ns (relative error: 4.9\% before rounding to ns) for the second eigenvector 
and 179 ps (relativ error: 5.3\%) for the third eigenvector. 
This shows that the dynamical reweighting method also works, when the perturbation acts on degrees of freedom which are part of the relevant coordinates on which the MSM is constructed.
%

\begin{figure*}[h!]
  \begin{center}
  \includegraphics[scale=1]{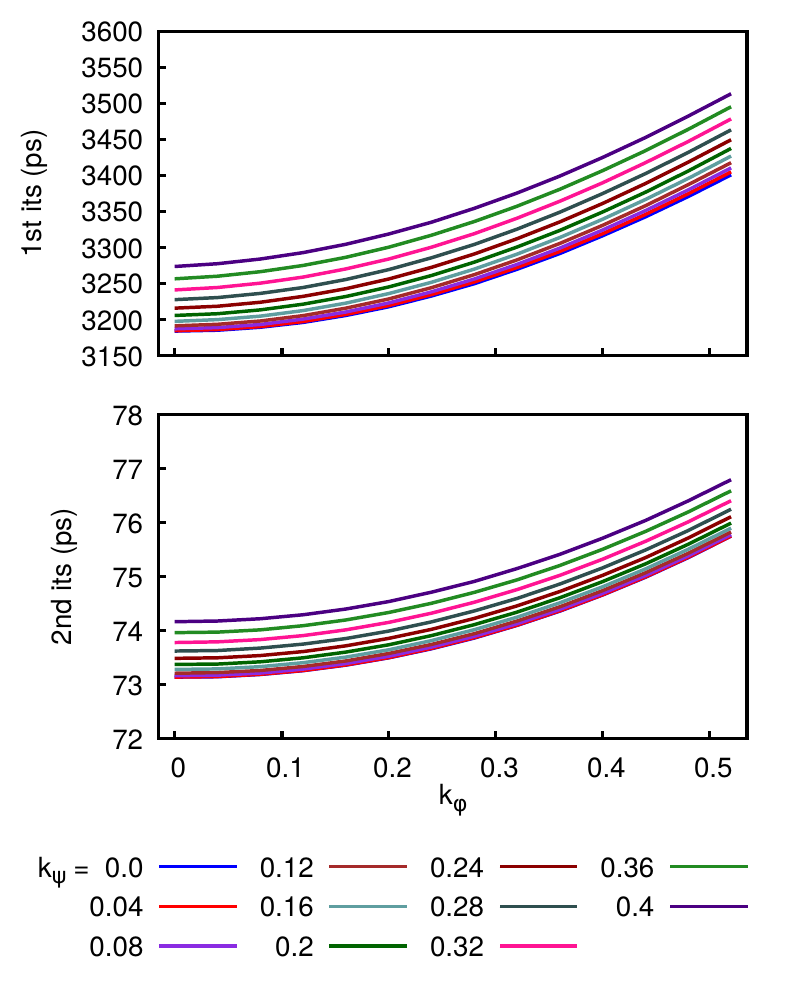}
    \caption{First and second implied time scales of alanine dipeptide in explicit solvent as function of the parameters $k_\phi, \ k_\psi$, estimated with the Girsanov reweighting method.}
    \label{graph:timescales_ala_wat}
  \end{center}
\end{figure*}

Fig.~\ref{graph:timescales_ala_wat} illustrates how to use the dynamical reweighting method to study the influence of a force constant on the molecular dynamics.
It shows the implied timescale of the second and third MSM eigenvectors of alanine dipeptide in explicit water as a function of the force constant $\kappa_{\phi}$, 
where the perturbation potential energy is given by eqs.~\ref{eq:peturbation_phi} and \ref{eq:perturbation_alanine}.
The scan has been repeated with different values of the force constant $\kappa_{\psi}$ for the potential energy function of the $\psi$-backbone dihedral angle
(eqs.~\ref{eq:peturbation_psi} and \ref{eq:perturbation_alanine}).
The dynamics is more sensitive to a change in the value of $\kappa_{\phi}$ than to a change of $\kappa_{\psi}$, but the overall effect of the perturbation is moderate.
It is important to point out that MSMs which are summarized in Fig.~\ref{graph:timescales_ala_wat} have been constructed from a single simulation at the reference potential energy function.
During this simulation, the It$\hat{\mbox{o}}$ integral $I(a)$ and the Riemann integral $R(a)$ have been calculated for $\kappa_{\phi}=1$ and $\kappa_{\psi}=1$, 
and the force constants have been scaled after the simulation during the construction of the MSM, as described in section \ref{sec:implementation}.

\section{Conclusion}
We have presented the Girsanov reweighting scheme, which is a method to study the dynamics of a molecular system subject to an (external) perturbation $U(\kappa, \mathbf{x})$ of the (reference) potential energy function $V(\mathbf{x})$.
It allows for the estimation of a dynamical model, e.g. a MSM, of the perturbed system from a simulation at the reference potential energy function.
The underlying assumption is that the equation of motion generates a path ensemble and that we can define a probability measure on this ensemble.
A perturbation of the potential energy function causes a modification of the probability measure. 
The Girsanov theorem guarantees that the probability ratio between these two measures exists (under certain conditions)
and leads to an analytical expression for this ratio (eq.~\ref{eq:reweightPathSpace01}).
By reformulating the MSM transition probabilities as path ensemble averages, we can apply the Girsanov reweighting scheme to obtain the transition probabilities of the perturbed system 
from a set of paths generated at the reference potential energy function.
The method can be extended to the variational approaches \cite{Nuske2014, Vitalini2015b}, milestoning approaches \cite{Schuette2011, Lemke2016} or tensor approaches  \cite{Nuske2016} to molecular dynamics, since in each of these methods the molecular transfer operator is discretized and the resulting matrix elements are estimated
as path ensemble averages.

Calculating the path probability ratio requires the knowledge of the random forces at each integration time step. 
For the explicit solvent simulations, we have introduced stochastic forces by using a Langevin thermostat.
In an efficient implementation of the method, two terms which are needed to calculate the probability ratio should be calculated ``on the fly'' during the simulation. We have demonstrated this using the MD simulation toolkit OPENMM \cite{b7}.
Two other dynamical reweighting schemes for MSMs have been published in recent years. 
In the reweighting scheme for parallel tempering simulations \cite{Prinz2011c, Chodera2011}, the path probability density is defined for time discretized paths at a reference temperature and then reweighted to different temperatures using statistically optimal estimators \cite{Shirts2008, Minh2009}.
Like the Girsanov reweighting scheme, this method relies on the random forces at each integration time step and reweights the contribution of each path to the estimate of the transition probability individually.
We remark that the reweighting to different thermodynamic states cannot be extended to the limiting case of continuous paths because for different volatilities the path probability ratio relative to the Wiener process cannot  be defined (see appendix \ref{sec:continuousPaths}). 
This however seems to be of little practical importance. 
In the TRAM method \cite{Mey2014, Wu2014, Wu2016}, rather than reweighting the probabilities of each individual path, the MSM transition probabilities $T_{ij}(\tau)$ are directly reweighted to a different potential energy function or a different thermodynamic state using a maximum likelihood estimator for the transition counts.
This becomes possible, if one additionally assumes that the dynamics is in local equilibrium within each microstate of the MSM, 
which possibly renders the method more sensitive to the MSM discretization than path based reweighting methods.

We have tested the Girsanov reweighting method on several systems, ranging from diffusion in a two-dimensional potential energy surface to alanine dipeptide and valine dipetide in implicit and explict water.
Importantly, the direct simulations of the perturbed potential energy function (Figs.~\ref{graph:manybody}, \ref{graph:alanine}, \ref{graph:valine})
are only included as a validation for the method. 
In an actual application one would only simulate the system at the reference potential energy function and then reweight to the perturbed potential energy function, 
thus saving the computational time of the direct simulation of the perturbed system.
Girsanov reweighting could be useful in several areas of research. 
First, one can very efficiently test the influence of a change in the potential energy function on the dynamics of the molecule. 
The influence of a change in the force constant on the dynamics is particularly easy to study.
The Girsanov reweighting method can therefore be applied to improve the dynamical properties of force fields \cite{Vitalini2015}, 
by for example tuning the force constants to match an experimentally measured correlation time.
Similarly, one can use Girsanov reweighting to understand the influence of restraining potentials \cite{Cesari2016, Keller2007} on the dynamics of the system.
Second, the method can be used to understand which degrees of freedom have the largest influence on the slow modes of the molecule
\cite{Tsourtis2015, Arampatzisa2016}.
For example, for alanine dipeptide, we showed that the slow dynamic modes are more sensitive to a force field variation in the $\phi$-backbone dihedral angle
than they are to a variation in the $\psi$-backbone dihedral angle.
Last but not least, Girsanov reweighting can be used to account for the effect of any external potential which has been added to the simulation in order to enhance the sampling.
Thus, one can for example estimate MSMs from metadynamics simulations\cite{Huber1994, Laio2002}, Hamilton replica exchange simulations  \cite{Sugita2000} or umbrella sampling simulations \cite{Torrie1977}.

\section{Supplementary material}
Example script for the implementation of the dynamical reweigthing method with OpenMM \cite{b7}.

\begin{acknowledgments}
This research has been  funded by Deutsche Forschungsgemeinschaft (DFG) through grant CRC 1114 ``Scaling Cascades in Complex Systems'', 
Project B05 ``Origin of the scaling cascades in protein dynamics''.
\end{acknowledgments}

\appendix

\section{Discussion over path probability measure of time-discrete and continuos paths}
\label{sec:continuousPaths}
In the theory section, we assumed to work with time-discrete trajectories and we have defined the corresponding concepts of path ensemble and path probability measures. In this section, we show why we cannot extend these concepts to continuous paths.

Let's consider a diffusion process $x_t\in \Gamma \subset \mathbb{R}$ solution of the stochastic differential equation 
\begin{equation}
\mathrm{d}x_t = a(x_t) \mathrm{d}t + \sigma \mathrm{d}B_t, \ \ \ 0 \leq t \leq \tau \\
\label{eq:app1}
\end{equation}
where $a(\cdot)$ is a drift, $B_t$ is a Brownian motion, $\sigma$ is a volatility and $\tau$ is the total time. The equation can be discretized in time according to several schemes. We consider the Euler-Maruyana method:
\begin{equation}
x_{k+1} = x_k + a_k\Delta t + \eta_t \sigma \sqrt{\Delta t} 
\label{eq:app2}
\end{equation}
where $a_k=a(x_t)$, $\Delta t$ is a time-step and $\eta$ is a random number drawn from a standard Gaussian distribution.

We now recall the same definitions given in the theory section. A \emph{time-discrete path} is the set of the points $x_k$ generated by the equation \ref{eq:app2} $\omega=\lbrace x(t=0)=x_0,x_1,x_2,...,x_n\rbrace$
where $n \in \mathbb{N}$ is the total number of time steps and $\tau=n \Delta t$.

If we consider an ensemble of paths starting from the same point $x_0$ and length $\tau$, we can define a path space $\Omega_{\tau,x} = \Gamma^n$, a path probability measure $P$ and the associated path probability density $\mu_P(\omega)$ as defined respectively in eq. \eqref{eq:path_measure1} and eq. \eqref{eq:path_probabilityDensity1}

The path probability density function is defined as in the theory section, but here we remark that, in a rigorous formalism, a density function is a derivative respect to the Lebesgue measure $\mathrm{d}\omega = \mathrm{d}x_1 \mathrm{d}x_2 ... \mathrm{d}x_n$, i.e. $\mu_{P}(\omega) = \frac{\mathrm{d}P}{\mathrm{d}\omega}$.

For Brownian motion with drift, the transition probability density $p(x_{k-1},x_k;\Delta t)$  that appear in eq. \eqref{eq:path_measure1} and eq. \eqref{eq:path_probabilityDensity1} is well defined for time-discrete paths:
\begin{equation}
p(x_{k-1},x_k;\Delta t) =  \frac{1}{\sqrt{2\pi\Delta t \sigma^2}} \exp\left(-\frac{(x_k - x_{k-1} - a_k \Delta t )^2}{2\Delta t \sigma^2}\right)
	\label{eq:app5}
\end{equation}

Let's see now what happen if we take $\Delta t \rightarrow 0$, i.e. if we consider continuous paths:
\begin{enumerate}
\item The normalization constant tends to infinity:
\begin{equation}
\lim_{\Delta t \rightarrow 0}  \frac{1}{\sqrt{2\pi\Delta t \sigma^2}} = +\infty
\end{equation}
\item The quantity in the exponential function becomes a derivative of the process $x_t$:
\begin{equation}
\begin{aligned}
\frac{(x_k - x_{k-1} - a_k \Delta t )^2}{2\Delta t \sigma^2} 
= & \left(\frac{x_k - x_{k-1}}{\Delta t} - a_k\right)^2 \frac{\Delta t}{2 \sigma^2} \\
\end{aligned}
\end{equation}
Then, formally
\begin{equation}
\lim_{\Delta t \rightarrow 0} \left(\frac{x_k - x_{k-1}}{\Delta t} - a_k\right)^2 \frac{\Delta t}{2 \sigma^2} = \left(\frac{\mathrm{d}x_t}{\mathrm{d}t} - a(x_t)\right)^2 \frac{\mathrm{d} t}{2 \sigma^2}
\end{equation}
But the derivative $\frac{\mathrm{d}x_t}{\mathrm{d}t}$ is not defined, because even though Brownian motion is everywhere continuous, it is nowhere differentiable.
\item The continuous path is made by an infinite number of points. If $n \rightarrow +\infty$, then the infinitesimal volume space $\mathrm{d}\omega = \mathrm{d}x_1 \mathrm{d}x_2 ... \mathrm{d}x_\infty$ becomes infinite-dimensional. However, the Lebesgue measure of an infinite-dimensional space does not exist.
\end{enumerate}
In conclusion, we cannot define a path probability density for a continuous path respect to the Lebesgue measure. 

On the other hand we can define the probability density of a path $\omega$ respect to a Wiener measure $W$. By computing explicitly the ratio between the probability density associated to a (discrete) brownian motion with drift and a probability density associated to a Brownian motion without drift, but with the same volatility, we get the following expression:
\begin{equation}
\frac{\mu_{P}(\omega)}{\mu_W(\omega)} = \frac{\mathrm{d}P}{\mathrm{d}\omega} \frac{\mathrm{d}\omega}{\mathrm{d}W} = \exp \left(\sum_{k=0}^n \frac{  (x_k - x_{k-1}) a_k }{\sigma^2} \right) \exp \left(-\sum_{k=0}^n \frac{ a_k^2 \Delta t}{2 \sigma^2} \right)
	\label{eq:app6}
\end{equation}
We observe that if we take the limit $\Delta t \rightarrow 0$, the three problems described above are solved: 
\begin{enumerate}
\item The normalization constant cancels.
\item The term $x_k - x_{k-1}$ is not divided by the time-step $\Delta t$, so there is not derivative respect to time.
\item The ratio between the two probability densities cancels also the infinitesimal volume $\mathrm{d}\omega$. The new probability density is defined respect to the Wiener measure, not respect the Lebesgue measure.
\end{enumerate}
Thus in the continuous case the likelihood \eqref{eq:app6} is written as:
\begin{equation}
\lim_{\Delta t \rightarrow 0} \frac{\mu_{P}(\omega)}{\mu_W(\omega)} = \exp \left(\int_0^T \frac{  a(x_s) }{\sigma} \mathrm{d}B_s  -\frac{1}{2} \int_0^T \frac{ a(x_s)^2 \mathrm{d} s}{\sigma^2} \right)
	\label{eq:app7}
\end{equation}
This quantity denotes the probability density, in the continuos case, of a path $x_t$ solution of a diffusion process, respect to the Wiener measure. The complete computation of the formulas \eqref{eq:app6} and \eqref{eq:app7} is reported in the appendix C. The Girsanov theorem states the condition under which the likelihood \eqref{eq:app7} exists.

\section{The Girsanov theorem}
\label{sec:GirsanovTheorem}
The Girsanov theorem \citep{Girsanov1960,Oksendal2003} is an important result of stochastic analysis that states the conditions under which a change of measure is possible for diffusion processes. 

Let's consider two diffusion processes $x_t \in \mathbb{R}$ and $y_t \in \mathbb{R}$ defined on the probability space $(\Omega, \mathcal{F}, \lbrace \mathcal{F}_t \rbrace_{t\geq 0})$, respectively with probability measures $P$ and $Q$ on the filtration  $\lbrace \mathcal{F}_t \rbrace_{t\geq 0})$, be solutions of the stochastic differential equations:
\begin{equation}
\begin{aligned}
\mathrm{d}x_t = a(x_t) \mathrm{d}t + \sigma \mathrm{d}B_t, \ \ \ 0 \leq t \leq \tau \\
\mathrm{d}y_t = b(y_t) \mathrm{d}t + \sigma \mathrm{d}B_t, \ \ \ 0 \leq t \leq \tau
\end{aligned}
\end{equation}  
where the functions $a: \mathbb{R}^n \rightarrow \mathbb{R}^n$ and $b: \mathbb{R}^n \rightarrow \mathbb{R}^n$ are measurable, $\sigma$ is the volatility, $B_t$ is an $n$-dimensional Brownian motion, $\tau \leq \infty$ and $x_0 = y_0 \in \mathbb{R}^n$.

Suppose that there exists a process $\xi_t$ such that
\begin{equation}
\xi_t = -\frac{a(y_t)-b(y_t)}{\sigma}
\end{equation}
and define
\begin{equation}
M_\tau=\exp\left( -\int_0^\tau \xi_s \mathrm{d}B_s -\frac{1}{2}\int_0^\tau \xi_s^2\mathrm{d}s \right)
\end{equation}
If the Novikov condition 
\begin{equation}\label{eq:novikov}
\mathbb{E} \left[ \exp \left( \frac{1}{2} \int_0^\tau \xi_s^2\mathrm{d}s \right) \right]<\infty
\end{equation} is satisfied, then the process $M_\tau$ is a martingale [\citep{Oksendal2003},pag.31] and we can define the probability measure $Q$ on $\mathcal{F}_t$ by the Radon-Nikodym derivative
\begin{equation}\label{eq:radon}
M_\tau=\frac{\mathrm{d}Q}{\mathrm{d}P}
\end{equation}
and the process
\begin{equation}
\mathrm{d} \hat B_t = \xi_t \mathrm{d}t + \mathrm{d} B_t
\end{equation}
is a new Brownian motion w.r.t. $Q$.
As consequence, for any measurable functional $f(x)$,
\begin{equation}\label{eq:gir_final}
\mathbb{E}_P \left[f(x_1,...,x_n) \right] = \mathbb{E}_Q \left[f(y_1,...,y_n)\right] = \mathbb{E}_P \left[M_\tau f(y_1,...,y_n) \right]
\end{equation}
\paragraph*{Remark} 
The measures $P$ and $Q$ on the filtration $\mathcal{F}_t$ represent the probability measures on a path space. In other words, given a set of trajectories, $P$ and $Q$ return the probability of a certain path. The concept of absolute continuity means that if a path is not allowed under $P$, then it is not allowed under $Q$:
\begin{equation}
\label{eq:abscon}
P(x_1\in F_1,...,x_n \in F_n)=0 \ \Rightarrow \ Q(x_1\in F_1,...,x_n \in F_n)=0 
\end{equation}
for every measurable set $F_1,...,F_n \subseteq \mathcal{F}_\tau$ with $\tau>0$ fixed. Because $M_\tau$ is strictly positive, the probability measure $P$ is also absolutely continuous w.r.t $Q$:
\begin{equation}
\label{eq:abscon1}
Q(x_1\in F_1,...,x_n \in F_n)=0 \ \Rightarrow \ P(x_1\in F_1,...,x_n \in F_n)=0 
\end{equation}
then the two measures $P$ and $Q$ are equivalent (or mutually absolutely continuous w.r.t each other) and we can write:
\begin{equation}
\label{eq:abscon2}
P(x_1\in F_1,...,x_n \in F_n)>0\ \Longleftrightarrow \ Q(x_1\in F_1,...,x_n \in F_n)>0 
\end{equation}

\section{Derivation of the Girsanov formula}
\label{sec:GirsanovFormula}
Let's now consider two diffusion processes with different drift terms, solutions of the stochastic differential equations:
\begin{gather}\label{eq:sde3}
\mathrm{d}x_t = a(x_t) \, \mathrm{d}t + \sigma \mathrm{d}B_t  \\
\label{eq:sde4}
\mathrm{d}x_t = b(x_t) \, \mathrm{d}t + \sigma \mathrm{d}B_t    
\end{gather}
with initial conditions $x_0=x \in \mathbb{R}$.
We discretize the equations \ref{eq:sde3}, \ref{eq:sde4} with the Euler-Maruyama method:
\begin{gather} 
x_{k+1} = x_k + a_k\Delta t + \eta \sigma \sqrt{\Delta t} \\
x_{k+1} = x_k + b_k\Delta t + \eta \sigma \sqrt{\Delta t} 
\end{gather}
where $a_k= a(x_k)$ and $b_k = b(x_k)$ are two different drift terms.
The transition probability densities are defined respectively as
\begin{equation}
p(x_{k-1},x_k;\Delta t) =  \frac{1}{\sqrt{2\pi\Delta t \sigma^2}} \exp\left(-\frac{(x_k - x_{k-1} - a_k \Delta t)^2}{2\Delta t \sigma^2}\right)
\end{equation}
for the first equation \ref{eq:sde3} and
\begin{equation}
p(x_{k-1},x_k;\Delta t) =  \frac{1}{\sqrt{2\pi\Delta t \sigma^2}} \exp\left(-\frac{(x_k - x_{k-1} - b_k \Delta t )^2}{2\Delta t \sigma^2}\right)
\end{equation}
for the second equation \ref{eq:sde4}.

The path probability density $\mu_{P_a}(\omega)$ associated to the equation \ref{eq:sde3} and $\mu_{P_b}(\omega)$ associated to the equation \ref{eq:sde4} are defined as in eq. \ref{eq:path_probabilityDensity1}. Thus, the ratio is given as
\begin{equation}
\begin{aligned}
L = &\frac{\mu_{P_b}(x_0,...,x_n)}{\mu_{P_a}(x_0,...,x_n)} \\
= & \frac{\exp \left(-\sum_{k=0}^n \frac{(x_k - x_{k-1} - b_k\Delta t)^2}{2\Delta t \sigma^2} \right)}{\exp \left(-\sum_{k=0}^n \frac{(x_k - x_{k-1} - a_k\Delta t)^2}{2\Delta t \sigma^2} \right)} \\
= & \frac{\exp \left(-\sum_{k=0}^n \frac{x_k^2 + x_{k-1}^2 + b_k^2 \Delta t^2 - 2 x_k x_{k-1} - 2x_k b_k \Delta t + 2 x_{k-1} b_k\Delta t}{2\Delta t \sigma^2} \right)}{\exp \left(-\sum_{k=0}^n \frac{x_k^2 + x_{k-1}^2 + a_k^2 \Delta t^2 - 2 x_k x_{k-1} - 2x_k a_k \Delta t + 2 x_{k-1} a_k\Delta t}{2\Delta t \sigma^2} \right)} \\
= & \frac{\exp \left(-\sum_{k=0}^n \frac{(x_k - x_{k-1} )^2}{2\Delta t \sigma^2} \right) \exp \left(-\sum_{k=0}^n \frac{ b_k^2 \Delta t^2  - 2x_k b_k \Delta t + 2 x_{k-1} b_k\Delta t}{2\Delta t \sigma^2} \right)}{\exp \left(-\sum_{k=0}^n \frac{(x_k - x_{k-1} )^2}{2\Delta t \sigma^2} \right) \exp \left(-\sum_{k=0}^n \frac{ a_k^2 \Delta t^2  - 2x_k a_k \Delta t + 2 x_{k-1} a_k\Delta t}{2\Delta t \sigma^2} \right)} \\
= & \exp \left(-\sum_{k=0}^n \frac{ \left( b_k^2 - a_k^2 \right) \Delta t^2  - 2x_k \left( b_k - a_k \right) \Delta t + 2 x_{k-1} \left( b_k - a_k \right) \Delta t}{2\Delta t \sigma^2} \right) \\
= & \exp \left(-\sum_{k=0}^n \frac{ - 2x_k \left( b_k - a_k\right) \Delta t + 2 x_{k-1} \left( b_k - a_k \right)\Delta t}{2\Delta t \sigma^2} \right) \exp \left(-\sum_{k=0}^n \frac{ \left( b_k^2 - a_k^2 \right) \Delta t^2}{2\Delta t \sigma^2} \right)\\
= & \exp \left(\sum_{k=0}^n \frac{  (x_k - x_{k-1}) \left( b_k - a_k \right) }{\sigma^2} \right) \exp \left(-\sum_{k=0}^n \frac{ \left( b_k^2 - a_k^2 \right) \Delta t}{2 \sigma^2} \right)
\end{aligned}
\end{equation}
If we take the limit $\Delta t \rightarrow 0$, the two sums in the exponential function converge respectively to the Ito integral and the Riemann integral. The jump $x_k - x_{k-1}$ is Wiener jump with drift $a_k \Delta t$:
\begin{equation}
\begin{aligned}
\sum_{k=0}^n (b_k-a_k)(x_k - x_{k-1}) \rightarrow & \int_0^T (b(x_s)-a(x_s)) \mathrm{d}x_s \\ 
= & \int_0^T (b(x_s)-a(x_s)) \mathrm{d} \hat B_s \\
= & \int_0^T (b(x_s)-a(x_s))(a(x_s)\mathrm{d}s + \sigma \mathrm{d} B_s) \\
= & \int_0^T a(x_s)b(x_s)\mathrm{d}s + b(x_s) \sigma \mathrm{d} B_s - a(x_s)^2\mathrm{d}s - a(x_s) \sigma \mathrm{d} B_s
\end{aligned}
\end{equation}
Finally we have:
\begin{equation}
\begin{aligned}
L = & \exp \left(\int_0^T \frac{a(x_s)b(x_s)\mathrm{d}s + b(x_s) \sigma \mathrm{d} B_s - a(x_s)^2\mathrm{d}s - a(x_s) \sigma \mathrm{d} B_s}{\sigma^2} \right) \exp \left(-\frac{1}{2} \int_0^T \frac{b(x_s)^2 - a(x_s)^2}{\sigma^2} \mathrm{d} s  \right) \\
 = & \exp \left(\int_0^T \frac{b(x_s) \sigma \mathrm{d} B_s - a(x_s) \sigma \mathrm{d} B_s }{\sigma^2} \right) \exp \left(\int_0^T \frac{a(x_s)b(x_s) - a(x_s)^2}{\sigma^2}\mathrm{d}s-\frac{1}{2} \int_0^T \frac{b(x_s)^2 - a(x_s)^2}{\sigma^2} \mathrm{d} s  \right) \\
 = & \exp \left(\int_0^T \frac{b(x_s) - a(x_s)} \sigma \mathrm{d} B_s \right) \exp \left(-\frac{1}{2} \int_0^T \frac{b(x_s)^2 + a(x_s)^2 - 2a(x_s)b(x_s)}{\sigma^2} \mathrm{d} s  \right) \\
  = & \exp \left(\int_0^T \frac{b(x_s) - a(x_s)} \sigma \mathrm{d} B_s \right) \exp \left(-\frac{1}{2} \int_0^T \frac{(b(x_s)-a(x_s))^2}{\sigma^2} \mathrm{d} s  \right) \\
\end{aligned}
\end{equation}

Thus we get the more general Girsanov formula valid also for two diffusion processes with different drifts.
\[
L = \exp \left(\int_0^T \frac{b(x_s) - a(x_s)} \sigma \mathrm{d} B_s -\frac{1}{2} \int_0^T \left(\frac{b(x_s)-a(x_s)}{\sigma}\right)^2 \mathrm{d} s  \right)
\]


\end{document}